%% file: manuscript.tex
\documentclass[useAMS,usenatbib,referee]{biom}

\input{preamble.tex}

\title[Competing Risk Modeling with Bivariate Varying Coefficients]{Understanding the dynamic impact of COVID-19 through competing risk modeling with bivariate varying coefficients}
\author{Wenbo Wu$^{1}$, John D. Kalbfleisch$^{2,3}$, Jeremy M. G. Taylor$^{2}$, Jian Kang$^{2,3}$, and Kevin He$^{2,3,*}$\email{\href{mailto:kevinhe@umich.edu}{kevinhe@umich.edu}} \\
$^{1}$Departments of Population Health and Medicine \\
Center for Surgical and Transplant Applied Research \\
New York University Grossman School of Medicine, New York, New York 10016, U.S.A. \\
$^{2}$Department of Biostatistics \\
University of Michigan, Ann Arbor, Michigan 48109, U.S.A. \\
$^{3}$Kidney Epidemiology and Cost Center \\
University of Michigan, Ann Arbor, Michigan 48109, U.S.A.}

\begin{document}

%  This will produce the submission and review information that appears
%  right after the reference section.  Of course, it will be unknown when
%  you submit your paper, so you can either leave this out or put in 
%  sample dates (these will have no effect on the fate of your paper in the
%  review process!)

% \date{{\it Received Month} 202x. {\it Revised Month} 202x.  {\it
% Accepted Month} 202x.}

\pagerange{\pageref{firstpage}--\pageref{lastpage}} 
% \volume{xx}
% \pubyear{202x}
% \artmonth{Month}

%\doi{10.1111/biom.xxxxx}

\label{firstpage}

% put the summary for your paper here

\begin{abstract}
The coronavirus disease 2019 (COVID-19) pandemic has exerted a profound impact on patients with end-stage renal disease relying on kidney dialysis to sustain their lives. Motivated by a request by the U.S. Centers for Medicare \& Medicaid Services, our analysis of their postdischarge hospital readmissions and deaths in 2020 revealed that the COVID-19 effect has varied significantly with postdischarge time and time since the onset of the pandemic. However, the complex dynamics of the COVID-19 effect trajectories cannot be characterized by existing varying coefficient models. To address this issue, we propose a bivariate varying coefficient model for competing risks within a cause-specific hazard framework, where tensor-product B-splines are used to estimate the surface of the COVID-19 effect. An efficient proximal Newton algorithm is developed to facilitate the fitting of the new model to the massive Medicare data for dialysis patients. Difference-based anisotropic penalization is introduced to mitigate model overfitting and the wiggliness of the estimated trajectories; various cross-validation methods are considered in the determination of optimal tuning parameters. Hypothesis testing procedures are designed to examine whether the COVID-19 effect varies significantly with postdischarge time and the time since pandemic onset, either jointly or separately. Simulation experiments are conducted to evaluate the estimation accuracy, type I error rate, statistical power, and model selection procedures. Applications to Medicare dialysis patients demonstrate the real-world performance of the proposed methods.
\end{abstract}

\begin{keywords}
cause-specific hazard, COVID-19, cross-validation, dialysis patients, difference-based anisotropic penalization, tensor-product B-splines
\end{keywords}

\maketitle

\section{Introduction}
\label{s:intro}
This paper grows out of our investigation in response to the request by the U.S. Centers for Medicare \& Medicaid Services (CMS) on the influence of the coronavirus disease 2019 (COVID-19) pandemic on patients with end-stage renal disease (ESRD) \citep[][]{wu2021covid19}. Our goal is to inform evidence-based COVID-19 adjustment in the implementation of ESRD quality measures, especially for postdischarge patient outcomes. These quality measures have been routinely reported on Care Compare--Dialysis Facilities \citep{medicaredfc2021} to assess dialysis facilities in the ESRD Quality Incentive Program \citep{esrdqip2021}. The calculation of pandemic-adjusted ESRD quality metrics largely depends on how COVID-19 as a risk factor should be accounted for in statistical modeling; any switch in measure-based flagging (e.g., from average to worse than expected) resulting from COVID-19 adjustment would lead to a substantial change in performance-based payments to dialysis facilities. This significant consequence indicates the high-stakes nature of our statistical endeavors.

\begin{figure}
    \centering
    \begin{subfigure}[t]{0.49\textwidth}
        \centering
        \includegraphics[width=\linewidth]{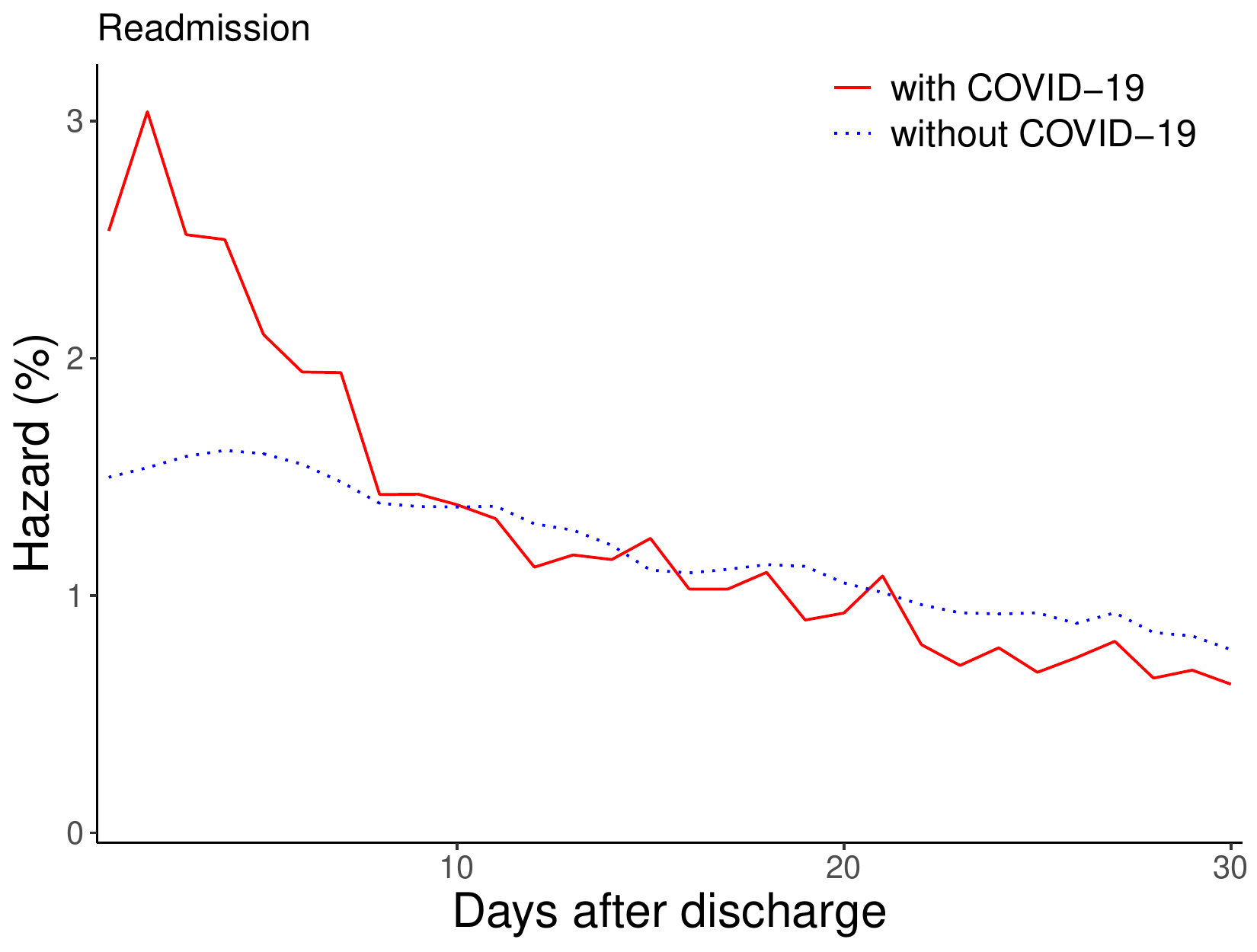} 
        \caption{hazard of readmission} \label{fig:UHR_pd_haz}
    \end{subfigure}
    \begin{subfigure}[t]{0.49\textwidth}
        \centering
        \includegraphics[width=\linewidth]{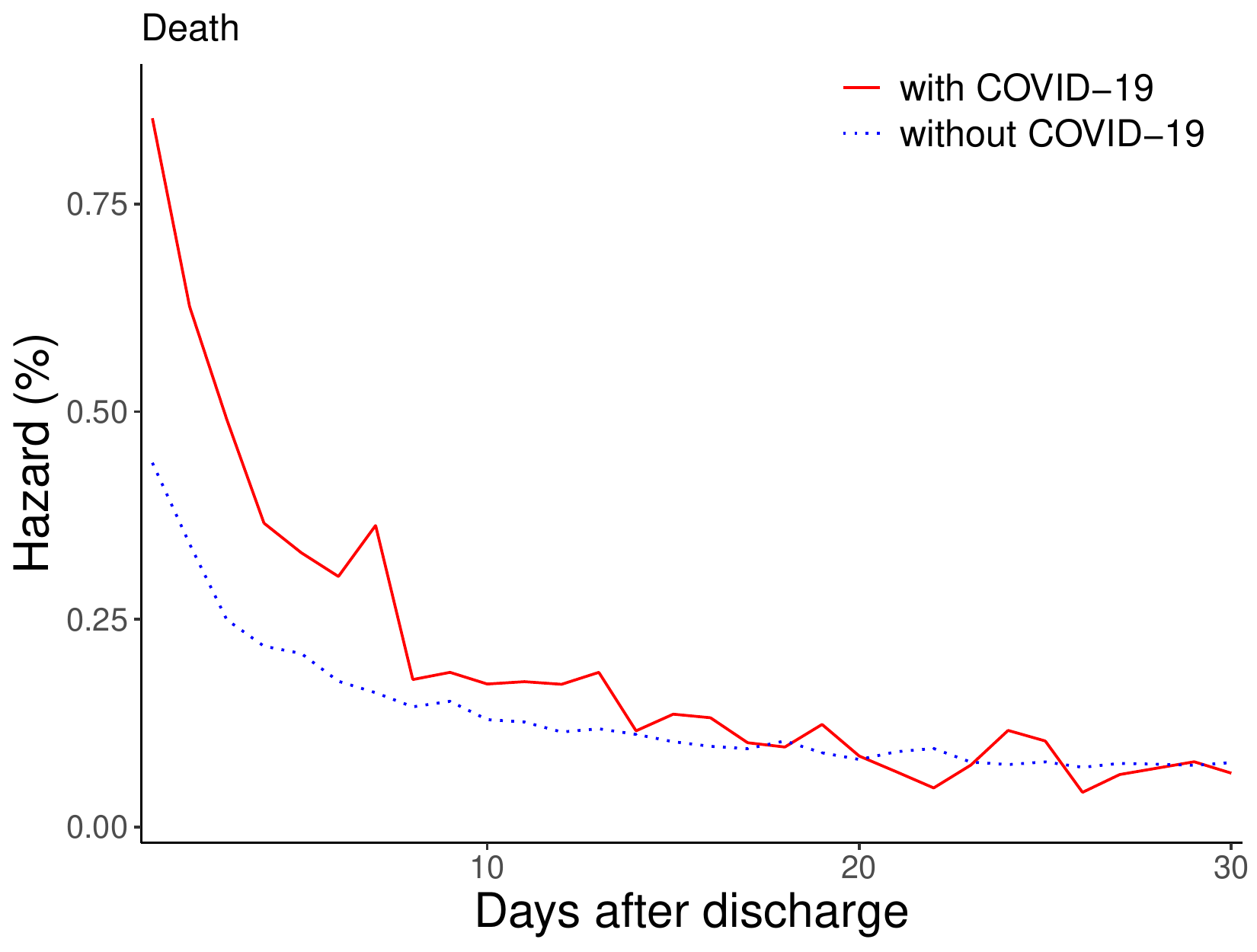} 
        \caption{hazard of death} \label{fig:death_pd_haz}
    \end{subfigure}
    
    \begin{subfigure}[t]{0.49\textwidth}
    \centering
        \includegraphics[width=\linewidth]{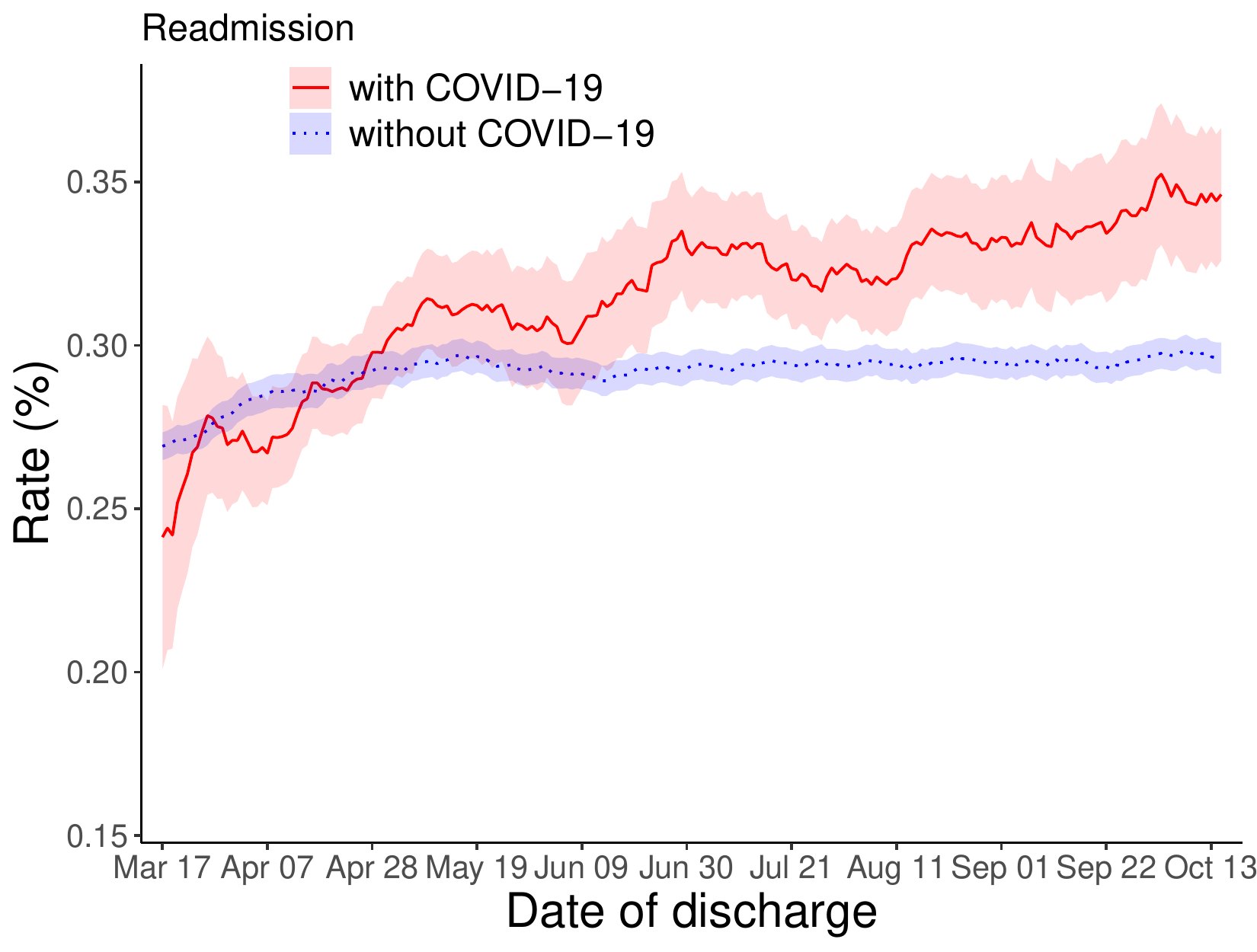}
        \caption{rate of readmission} \label{fig:UHR_cal_rate}
    \end{subfigure}
    \begin{subfigure}[t]{0.49\textwidth}
    \centering
        \includegraphics[width=\linewidth]{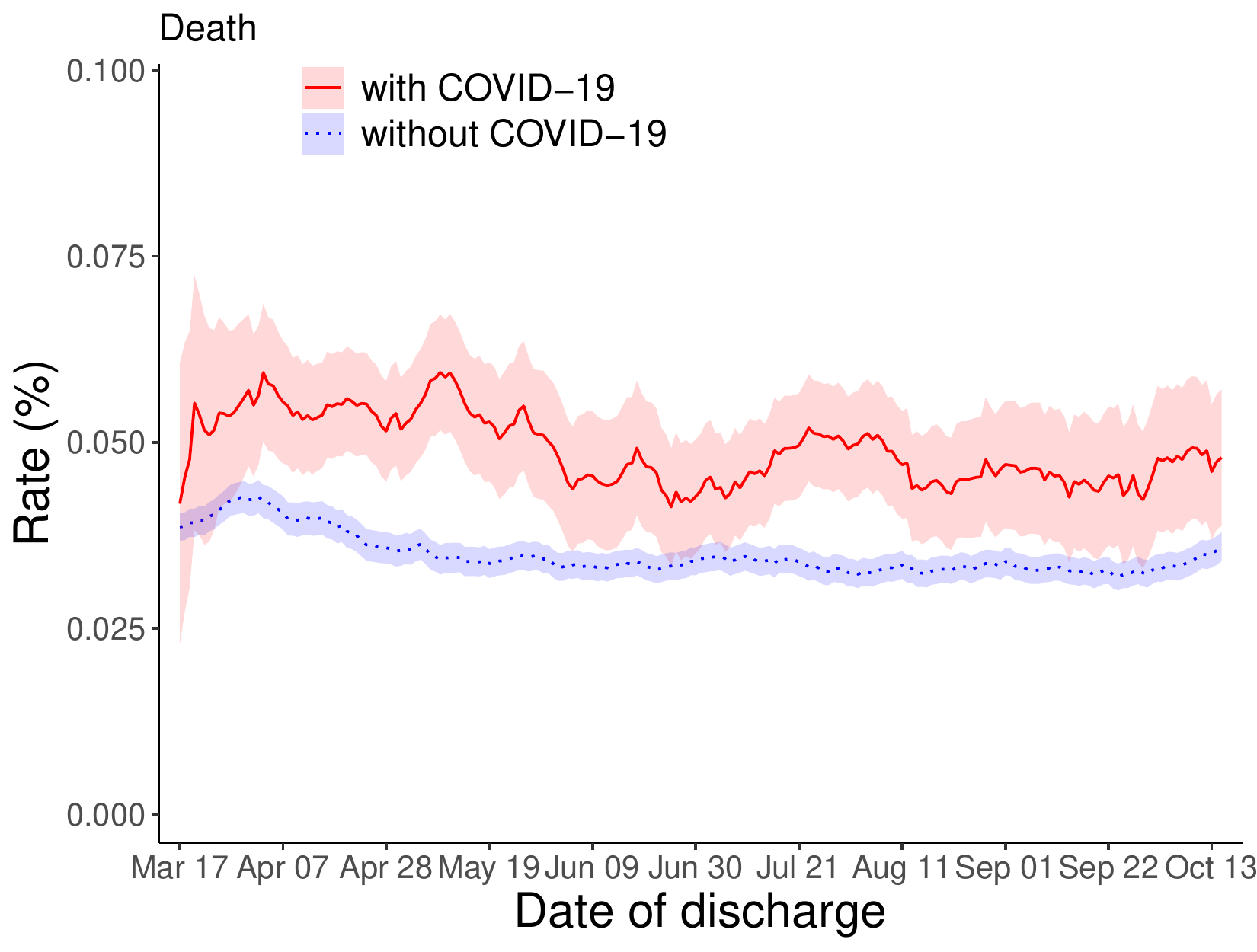} 
        \caption{rate of death} \label{fig:death_cal_rate}
    \end{subfigure}
    \caption{Panels (a) and (b) present unadjusted cause-specific hazard curves of unplanned hospital readmission and death, respectively, from January 1, 2020 to October 31, 2020. On each postdischarge day, the unadjusted hazard of readmission or death was defined as the number of readmissions or deaths occurring over that day divided by the number of discharges at risk for readmission (or death) at the beginning of that day. Panels (c) and (d) present rates of unplanned hospital readmission and death, respectively, among discharges with and without in-hospital COVID-19 from March 17, 2020 to October 15, 2020. Monthly rates and their 95\% confidence intervals were calculated on a rolling basis. \label{fig:covid_postdischarge}}
\end{figure}

To understand the impact of COVID-19 on patients requiring routine kidney dialysis for appropriate risk adjustment in CMS reporting, we explored their postdischarge readmissions and deaths by in-hospital COVID-19 diagnosis (with versus without COVID-19). Included in the data were 436,745 live hospital discharges of 222,154 Medicare dialysis beneficiaries from 7,871 dialysis facilities throughout the first ten months of 2020. The top two panels of \Cref{fig:covid_postdischarge} shows that within a week of hospital discharge, the descending (unadjusted) cause-specific hazard curves of readmission and death were substantially higher for the group with COVID-19 than the group without. \Cref{fig:UHR_cal_rate} shows that the rate of readmission increased in both groups between mid-March and mid-May; from early June onward, the rate of readmission among discharges with COVID-19 began to significantly surpass the rate of readmission among discharges without. \Cref{fig:death_cal_rate} indicates that the rate of death in both groups started at a relatively high level and then overall decreased until mid-October; the rate of death among discharges with COVID-19 remained significantly higher than the rate among discharges without throughout the ten months. Despite the fact that other risk factors were not adjusted for, these preliminary findings indicate that the impact of COVID-19 was constantly changing with both postdischarge (\Cref{fig:UHR_pd_haz,fig:death_pd_haz}) and calendar time (\Cref{fig:UHR_cal_rate,fig:death_cal_rate}).

Existing risk adjustment models for quality measure development and health care provider monitoring mostly treat the outcome of interest as a binary variable, using logistic regression with fixed and random effects to indicate inter-provider variation \citep[e.g.,][]{he2013evaluating,kalbfleisch2013monitoring,estes2018time,estes2020profiling,wu2022improving,normand1997statistical,ohlssen2007hierarchical,ash2012statistical,mcgee2020outcome}. Because these models do not account for event timing, they cannot be applied to our COVID-19 study to discover the important evidence of postdischarge variation; a time-to-event modeling framework would better meet the analytical needs in this setting. In addition, the unusual dynamics of the COVID-19 effect calls for a distinctive varying coefficient model that provides a unified characterization of the significant variations with both postdischarge and calendar time, and a systematic inferential procedure testing the two-dimensional variations either jointly or separately. Unfortunately, such a flexible and comprehensive model is still lacking in the statistical literature.

Motivated by the pressing need for novel statistical methods to appropriately analyze the dynamic impact of COVID-19, we develop a spline-based bivariate varying coefficient model, treating postdischarge readmission and death as competing risks within a cause-specific hazard framework. Unlike existing time-varying coefficient models for time-to-event outcomes \citep[e.g.,][]{zucker1990nonparametric, gray1992flexible, hastie1993varying, verweij1995time, tutz2004flexible, he2017modeling, he2021stratified, wu2022scalable}, the proposed model formulates the effect of a risk factor (e.g., in-hospital COVID-19 diagnosis in our applications) as a bivariate function of both event time (e.g., postdischarge time to a readmission or death, hereafter postdischarge time) and an external covariate (e.g., calendar time since pandemic onset, hereafter calendar time). Tensor-product B-splines \citep[][\S12.2]{schumaker2007spline} are employed to estimate the surface of the bivariate COVID-19 effect, thereby allowing complex variation trajectories along two different dimensions. Although tensor-product B-splines were previously used to model interactions between two continuous risk factors \citep{gray1992flexible}, our study is the first to use this technique to characterize the complexly varying effect of a risk factor in a competing risk analysis.

Fitting the bivariate varying coefficient model to the massive postdischarge outcome data for Medicare dialysis patients poses significant computational issues that no existing methods can handle. Current methods rely on expanding a single observation into multiple records from the baseline until the observed time \citep{therneau2020using}. With large-scale data, this approach leads to prolonged convergence and overloaded memory even for univariate time-varying effect modeling \citep{wu2022scalable}, let alone bivariate varying coefficients. Moreover, the presence of extremely distributed binary covariates often introduce numerical instability with ill-conditioned Hessian matrices. To address these challenges, we develop a tensor-product proximal Newton algorithm that optimizes the unpenalized log-partial likelihood. This algorithm efficiently extends the approach by \citet{wu2022scalable} to a two-dimensional setting. Leveraging the property of B-splines, we propose a hypothesis testing framework with respect to both univariate and bivariate variation of the COVID-19 effect.

To mitigate model overfitting and the wiggliness of the estimated COVID-19 effect surface in a multivariate setting, we also introduce difference-based anisotropic penalization \citep{wood2000modelling, wood2006low, eilers2021practical} to the original log-partial likelihood, where the penalization is applied against the deviation from a constant coefficient model, and the degree of the penalty is regulated through dimension-specific sets of tuning parameters. The asymptotic distribution of the resulting penalized estimates is investigated under mild conditions, and a corresponding inference procedure that generalizes the test of \citet{gray1992flexible} is developed. To determine optimal tuning parameters, we evaluate various methods of cross-validation and extend the method of cross-validated deviance residuals to the setting with varying coefficients.

The rest of this article is organized as follows: \Cref{s:model} introduces the bivariate varying coefficient model for competing risks. \Cref{s:estandinfunpen} presents estimation and inference methods based on the unpenalized partial likelihood. In \Cref{s:estandinfpen}, we develop estimation and inference methods based on the penalized partial likelihood. Next, we demonstrate and evaluate the proposed methods with two applications to Medicare dialysis patients in \Cref{s:RDA} and simulation experiments in \Cref{s:SDA}. \Cref{s:dis} concludes with a discussion.

\section{Model}
\label{s:model}
First, we present a competing risk model with bivariate varying coefficients. For the $i$th subject in the $g$th stratum ($g = 1, \ldots{}, G$, $i = 1, \ldots{}, n_g$, where $n_g$ denotes the total number of subjects in the $g$th stratum, i.e., dialysis facility in our applications), let $T_{gi}$, $C_{gi}$, and $X_{gi} \coloneqq \min\{T_{gi}, C_{gi}\}$ denote the failure, censoring and observed times, respectively. Let $\bZ_{gi}$ denote a vector of $p$ covariates associated with $p$ bivariate varying coefficients, and let $\bW_{gi}$ denote a vector of $q$ covariates with invariant coefficients. For ease of notation and due to the interest of our applications, we assume that all bivariate varying coefficients depend upon a single effect modifying covariate $\brX_{gi}$, although the dependence can be easily relaxed to be coefficient-specific. Further, let $J_{gi}$ be a random variable such that $J_{gi} = j$ ($j = 1, \ldots{}, m$) if subject $i$ in stratum $g$ has a failure of type $j$, and $J_{gi} = 0$ if that subject is censored. In our applications (details in \Cref{s:RDA}), $j$ indicates different postdischarge outcomes (unplanned hospital readmission and death) or discharge destinations (to home, to another health care facility, and in-hospital death or to hospice). Let $\Delta_{jgi} \coloneqq I(T_{gi} \leq C_{gi}, J_{gi}=j)$ indicate whether subject $i$ in stratum $g$ has a type $j$ failure, where $I(\cdot)$ is an indicator function, and let $\Delta_{gi} \coloneqq I(T_{gi} \leq C_{gi})$. We assume that conditional on $\bZ_{gi}$, $\bW_{gi}$ and $\brX_{gi}$, $T_{gi}$ and $C_{gi}$ are independent so that the censoring is non-informative.

We consider a stratified Cox relative risk model with semi-varying coefficients \citep{fan2008statistical}, i.e.,
\begin{equation}
\label{eq:cause-specifichazard}
\lambda_{jgi}(t \mid \bZ_{gi}, \bW_{gi}, \brX_{gi}) \coloneqq \lambda_{0jg}(t)\exp\left\{\bZ_{gi}^\top \bbeta_j(t, \brX_{gi}) + \bW_{gi}^\top \btheta_j\right\}, \quad j = 1, \ldots{}, m,
\end{equation}
where $\lambda_{jgi}(t \mid \bZ_{gi}, \bW_{gi}, \brX_{gi})$ denotes the stratum- and cause-specific hazard function for failure type $j$, $\lambda_{0jg}(t)$ denotes the baseline hazard function allowed to be arbitrary and assumed completely unrelated, $\bbeta_j(t, \brX_{gi}) \coloneqq [\beta_{j1}(t,\brX_{gi}), \ldots, \beta_{jp}(t,\brX_{gi})]^\top$ is a $p$-dimensional vector of varying coefficients, each of which is a bivariate function of time $t$ and covariate $\brX_{gi}$, and $\btheta_j$ is a $q$-dimensional vector of invariant coefficients. In our setting, $t$ denotes the time (in days) since hospital discharge or admission, and $\brX_{gi}$ denotes the discharge or admission time (in days) since the onset of the COVID-19 pandemic.

To approximate the surface of $\beta_{jl}(t, \brx)$, $l = 1, \ldots, p$, we span $\beta_{jl}(\cdot, \cdot)$ by tensor-product B-splines. Specifically,
\begin{equation}
\label{eq:bivartv}
\beta_{jl}(t, \brx)
\coloneqq
\brbB^\top(\brx) \bgamma_{jl} \bB(t)
= \sum_{\brk=1}^{\brK} \sum_{k=1}^K \gamma_{jl\brk k} \brB_{\brk}(\brx) B_k(t),
\end{equation}
where $\bB(t) \coloneqq [B_1(t), \ldots, B_K(t)]^\top$ and $\brbB(\brx) \coloneqq [\brB_1(\brx), \ldots, \brB_
{\brK}(\brx)]^\top$ are B-spline bases (with intercept terms) at $t$ and $\brx$, respectively, and $\bgamma_{jl} \coloneqq [\gamma_{jl\brk k}]$ is a $\brK \times K$ matrix of unknown control points for the $l$th bivariate varying coefficient $\beta_{jl}(\cdot, \cdot)$ of failure type $j$. The number $K$ (or $\brK$) of B-spline functions forming a basis $\bB(t)$ (or $\brbB(\brx)$) relates to the degree $d$ (or $\brd$) of the piecewise B-spline polynomials and to the number $u$ (or $\bru$) of interior knots in that $K = u + d + 1$ (or $\brK = \bru + \brd + 1$) \citep[][Theorem 4.4]{schumaker2007spline}. In reality, interior knots of the B-spline space $\bB(\cdot)$ can be chosen based on the quantiles of distinct failure times $\{X_{gi}: \Delta_{gi} = 1,\, i = 1, \ldots{}, n_g, g = 1, \ldots{}, G\}$ \citep{gray1992flexible,he2017modeling,he2021stratified,wu2022scalable}, and the interior knots of $\brbB(\cdot)$ can be set at the quantiles of covariates $\{\brX_{gi}: i = 1, \ldots{}, n_g, g = 1, \ldots{}, G\}$.

Note that \eqref{eq:bivartv} can be rewritten as
\begin{equation*}
\beta_{jl}(t, \brx) 
= \{\vec(\bgamma^\top_{jl})\}^\top \{\brbB(\brx) \otimes \bB(t)\},
\end{equation*}
where $\vec$ denotes the vectorization of a matrix, i.e., stacking columns of a matrix on top of one another, and $\otimes$ denotes the Kronecker product. It follows that 
\begin{equation*}
\label{eq:bivartvasvector}
\bbeta_j(t, \brx) = \bGamma_j \{\brbB(\brx) \otimes \bB(t)\},
\end{equation*}
where $\bGamma_j \coloneqq [\vec(\bgamma^\top_{j1}), \ldots, \vec(\bgamma^\top_{jp})]^\top$. Let $\bgamma_j \coloneqq \vec(\bGamma_j^\top)$, $\bgamma \coloneqq [\bgamma_1^\top, \ldots{}, \bgamma_m^\top]^\top$, and $\btheta \coloneqq [\btheta_1^\top, \ldots{}, \btheta_m^\top]^\top$. Given model \eqref{eq:cause-specifichazard}, we have the log-partial likelihood
\begin{equation}
\label{eq:logpartiallkd}
\ell(\bgamma, \btheta)
= \sum_{j=1}^m \ell_j (\bgamma_j, \btheta_j)
= \sum_{j=1}^m \sum_{g=1}^G \ell_{jg} (\bgamma_j, \btheta_j),
\end{equation}
in which
\begin{align}
\ell_{jg}(\bgamma_j, \btheta_j)
% &\coloneqq \sum_{i=1}^{n_g} \Delta_{jgi}\left [\bZ_{gi}^\top \bGamma_j \{\brbB(\brX_{gi}) \otimes \bB(X_{gi})\} + \bW_{gi}^\top \btheta_j \phantom{\left\{\sum_{r \in R_g(X_{gi})}\right\}}\right. \nonumber \\
% & \qquad -\log
% \left.\left\{\sum_{r \in R_g(X_{gi})} \exp \left(\bZ_{gr}^\top \bGamma_j \{\brbB(\brX_{gr}) \otimes \bB(X_{gi})\} + \bW_{gr}^\top \btheta_j \right) \right \} \right] \nonumber \\
&= \sum_{i=1}^{n_g} \Delta_{jgi} \left [\bL^\top_{gi}(X_{gi}) \bgamma_j + \bW_{gi}^\top \btheta_j
- \log
\left\{\sum_{r \in R_g(X_{gi})} \exp \left(\bL_{gr}^\top(X_{gi}) \bgamma_j + \bW_{gr}^\top \btheta_j\right)\right\} \right], \label{eq:cause-specificlogpartiallkd}   
\end{align}
$R_g(X_{gi}) \coloneqq \{r \in \{1, \ldots{}, n_g\}: X_{gr} \geq X_{gi}\}$ denotes the risk set of subject $i$ in stratum $g$, and $\bL_{gr}(X_i) \coloneqq \bZ_{gr} \otimes \brbB(\brX_{gr}) \otimes \bB(X_{gi})$. The gradient and Hessian matrix of $\ell_{jg}(\bgamma_j, \btheta_j)$ are available in Appendix A.

\section{Unpenalized partial likelihood approach}
\label{s:estandinfunpen}

\subsection{Estimation}
As noted before, the joint estimation of the bivariate varying coefficient functions $\beta_{jl}(\cdot, \cdot)$ and invariant coefficients $\btheta_j$ based on the unpenalized log-partial likelihood \eqref{eq:logpartiallkd} becomes computationally challenging, especially when the sample includes at least half a million subjects. To address this challenge, we develop a tensor product proximal Newton algorithm on the basis of \citet{wu2022scalable} to allow bivariate varying coefficient estimation. This approach is derived from the proximal operator \citep{parikh2014proximal} of the second-order Taylor approximation of the log-partial likelihood \eqref{eq:logpartiallkd}, leading to a modified Hessian matrix. The algorithm features accurate and efficient model fitting to large-scale competing risks data with millions of subjects and binary predictors of near-zero variance. Let $X_{jg1} < \cdots < X_{jgn_{jg}}$ denote the $n_{jg}$ distinct times of type $j$ failures within stratum $g$. For failure time $X_{jgb}$, $b = 1, \ldots{}, n_{jg}$, let $\bZ_{jgb}$, $\bW_{jgb}$, and $\brX_{jgb}$ denote $\bZ_{gi}$, $\bW_{gi}$, and $\brX_{gi}$, respectively, such that $\Delta_{jgi} = 1$ and $X_{gi} = X_{jgb}$. The algorithm is outlined as \Cref{alg:TPProxN}. For theoretical arguments justifying the convergence of the algorithm, the reader is referred to \citet{wu2022scalable}. In what follows, we will use carets to indicate unpenalized estimates resulting from this algorithm. For instance, $\hat\bgamma_{jl}$ denotes unpenalized estimates of $\bgamma_{jl}$.

\begin{algorithm}[htbp]
\setstretch{0.9}
\SetAlCapNameFnt{\footnotesize}
\SetAlCapFnt{\footnotesize}
\footnotesize
\SetAlgoLined
\SetKwRepeat{Do}{do}{while}
\For(\tcp*[f]{$m$ failure types}){$j \leftarrow 1$ \KwTo $m$}{
initialize $s \leftarrow 0$, $\lambda_0 > 0$, $\bgamma_j^{(0)} = \mathbf{0}$, and $\btheta_j^{(0)} = \mathbf{0}$\; 
set $\phi \in (0,0.5)$, $\psi \in (0.5, 1)$, $\delta \geq 1$ and $\epsilon > 0$\;
    \Do{$\eta^2 \geq 2\epsilon$}{
    \For(\tcp*[f]{$G$ distinct strata}){$g \leftarrow 1$ \KwTo $G$}{
        \For(\tcp*[f]{$n_{jg}$ distinct failure times}){$b \leftarrow 1$ \KwTo $n_{jg}$}{
            \For{$u \leftarrow 0$ \KwTo $2$}{
 $S^{(u)}_{jgb}(\bgamma_j^{(s)}, \btheta_j^{(s)}, X_{jgb}) = \sum_{r \in R_g(X_{jgb})}\exp\{\bL^\top_{gr}(X_{jgb}) \bgamma_j^{(s)} + \bW^\top_{gr} \btheta^{(s)}_j\}
\begin{bmatrix}
\bL_{gr}(X_{jgb}) \\
\bW_{gr}
\end{bmatrix}^{\odot u}$\;
            }
            \For{$w \leftarrow 1$ \KwTo $2$}{
                $\bU^{(w)}_{jgb}(\bgamma_j^{(s)}, \btheta_j^{(s)}, X_{jgb}) 
                = S^{(w)}_{jgb}(\bgamma_j^{(s)}, \btheta_j^{(s)}, X_{jgb}) / S^{(0)}_{jgb}(\bgamma_j^{(s)}, \btheta_j^{(s)}, X_{jgb})$\;
            }
            $\bV_{jgb}(\bgamma_j^{(s)}, \btheta_j^{(s)}, X_{jgb})
            = \bU^{(2)}_{jgb}(\bgamma_j^{(s)}, \btheta_j^{(s)}, X_{jgb}) - 
            \left[\bU^{(1)}_{jgb}(\bgamma_j^{(s)}, \btheta_j^{(s)}, X_{jgb})\right]^{\odot 2}$\;
        }
        }
        $\dot{\ell}_j(\bgamma_j^{(s)}, \btheta_j^{(s)})
        =\sum_{g=1}^G \sum_{q=1}^{n_j} \left \{
\begin{bmatrix}
\bL_{jgb}(X_{jgb}) \\
\bW_{jgb}
\end{bmatrix} - \bU^{(1)}_{jgb}(\bgamma_j^{(s)}, \btheta_j^{(s)}, X_{jgb})\right\}$\;
        $\ddot{\ell}_j(\bgamma_j^{(s)}, \btheta_j^{(s)})
        =-\sum_{g=1}^G \sum_{q=1}^{n_j} \bV_{jgb}(\bgamma_j^{(s)}, \btheta_j^{(s)}, X_{jgb})$\;
        $
\begin{bmatrix}
\Delta\bgamma_j^{(s)} \\
\Delta\btheta_j^{(s)}
\end{bmatrix}
        = \left[\bI / \lambda_s - \ddot{\ell}_j(\bgamma_j^{(s)}, \btheta_j^{(s)})/n \right]^{-1} \dot{\ell}_j(\bgamma_j^{(s)}, \btheta_j^{(s)})/n$ \tcp*[r]{Newton step}
         $\eta^2 = \dot{\ell}^\top_j(\bgamma_j^{(s)}, \btheta_j^{(s)})
\begin{bmatrix}
\Delta\bgamma_j^{(s)} \\
\Delta\btheta_j^{(s)}
\end{bmatrix}$ \tcp*[r]{$\eta$: Newton increment}
         $\nu \leftarrow 1$\;
        \lWhile(\tcp*[f]{line search}){$\sum_{g=1}^G\ell_{jg}(\bgamma_j^{(s)} + \nu\Delta\bgamma_j^{(s)}, \btheta_j^{(s)} + \nu\Delta\btheta_j^{(s)}) < \sum_{g=1}^G \ell_{jg}(\bgamma_j^{(s)}, \btheta_j^{(s)}) + \phi \nu \eta^2$}{$\nu \leftarrow \psi\nu$}
        $\bgamma_j^{(s+1)} = \bgamma_j^{(s)} + \nu\Delta\bgamma_j^{(s)}$\;
        $\btheta_j^{(s+1)} = \btheta_j^{(s)} + \nu\Delta\btheta_j^{(s)}$\;
        $\lambda_{s+1} = \delta \lambda_s$\;
        $s \leftarrow s + 1$\;
    }
}
\caption{Tensor Product Proximal Newton}
\label{alg:TPProxN}
\end{algorithm}

\subsection{Inference}
\label{s:inferunpen}
To examine the dynamic impact of COVID-19 among dialysis patients, it is logical to test whether a bivariate coefficient $\beta_{jl}(t, \brx)$ varies significantly with $t$ and $\brx$, either separately or jointly. By the property of B-splines, when $\gamma_{jl\brk k}$ remains constant with $k$, i.e., $\gamma_{jl\brk k} \equiv \gamma_{jl\brk \cdot}$ for any $\brk = 1, \ldots{}, \brK$, $\beta_{jl}(t, \brx)$ reduces to
\begin{equation*}
\label{eq:univartvcal}
\beta_{jl}(t, \brx)
= \sum_{\brk=1}^{\brK} \gamma_{jl\brk \cdot} \brB_{\brk}(\brx) \sum_{k=1}^K B_k(t)
= \sum_{\brk=1}^{\brK} \gamma_{jl\brk \cdot} \brB_{\brk}(\brx),
\end{equation*}
which no longer varies with $t$ due to the fact that $\sum_{k=1}^K B_k(t) = 1$. This relationship suggests the null hypothesis $H^{(t)}_{0}: \bC^{(t)} \vec(\bgamma^\top_{jl}) = \mathbf{0}$ for testing whether $\beta_{jl}(t, \brx)$ varies significantly with $t$, where
$\bC^{(t)} = \diag(\underbrace{\bD, \ldots{}, \bD}_{\brK})$ is a block diagonal matrix with $\brK$ diagonal blocks. Each of these blocks is a $(K-1) \times K$ first-order difference matrix $\bD$ of the form
\[
\begin{bmatrix}
1 & -1 & 0 & \cdots & 0 \\
0 & 1 & -1 & \cdots & 0 \\
\vdots & \vdots & \ddots & \ddots & \vdots \\
0 & 0 & \cdots & 1 & -1
\end{bmatrix}.
\]
A Wald test statistic associated with the null $H^{(t)}_{0}$ can thus be constructed as
\begin{equation}
\label{eq:Waldpdunpen}
\{\bC^{(t)}\vec(\hat\bgamma^\top_{jl})\}^\top \left[\bC^{(t)}\widehat\bM_{jl}\{\bC^{(t)}\}^\top\right]^{-1}\bC^{(t)}\vec(\hat\bgamma^\top_{jl}),    
\end{equation}
where $\widehat\bM_{jl}$ denotes the $l$th $K\brK \times K\brK$ diagonal block of $\{-\sum_{g=1}^G \ddot{\ell}_{jg}(\hat\bgamma_j, \hat\btheta_j)\}^{-1}$ with $\ddot{\ell}_{jg}(\bgamma_j, \btheta_j)$ being the Hessian matrix of $\ell_{jg}(\bgamma_j, \btheta_j)$. Under $H^{(t)}_{0}$, the test statistic approximately follows a chi-squared distribution with $\brK(K-1)$ degrees of freedom.

To test whether $\beta_{jl}(t, \brx)$ varies significantly with $\brx$, observe that when $\gamma_{jl\brk k} \equiv \gamma_{jl\cdot k}$ for any $k = 1, \ldots{}, K$, $\brk = 1, \ldots{}, \brK$, the bivariate coefficient
$
\beta_{jl}(t, \brx)
= \sum_{k=1}^K \gamma_{jl\cdot k} B_k(t)
$
no longer varies with $\brx$. The corresponding null hypothesis is $H^{(\brx)}_{0}: \bC^{(\brx)} \vec(\bgamma^\top_{jl}) = \mathbf{0}$ where $\bC^{(\brx)}$ is a $K(\brK-1) \times K\brK$ difference matrix of the $K$th order. The Wald test statistic is readily obtained by substituting $\bC^{(t)}$ in \eqref{eq:Waldpdunpen} with $\bC^{(\brx)}$. Similarly, the null hypothesis for testing whether $\beta_{jl}(t, \brx)$ varies significantly with both $t$ and $\brx$ is $H^{(t,\brx)}_{0}: \bC^{(t,\brx)} \vec(\bgamma^\top_{jl}) = \mathbf{0}$, where $\bC^{(t,\brx)}$ is a $(K\brK-1) \times K\brK$ first-order difference matrix. The Wald test statistic can be written by substituting $\bC^{(t)}$ in \eqref{eq:Waldpdunpen} with $\bC^{(t,\brx)}$.

\section{Penalized partial likelihood approach}
\label{s:estandinfpen}
% Penalization is useful in reducing the wiggliness of the estimated effect surface. In-depth accounts on the computational specifics of tensor-product B-splines can be found at \citet{eilers2003multivariate, eilers2021practical}. Anisotropic penalization was discussed in \citet{wood2000modelling,wood2006low} and \citet[\S4.3]{eilers2021practical}.

\subsection{Difference-based anisotropic penalization}
To mitigate overfitting and increase the smoothness of the estimated coefficient surface of $\beta_{jl}(\cdot,\cdot)$, we consider penalizing the column-wise and row-wise differences between adjacent control points of $\bgamma_{jl}$. The penalized log-partial likelihood can be written as
\[
\ell^{(\rP)}(\bgamma, \btheta)
= \sum_{j=1}^m \ell_j^{(\rP)}(\bgamma_j, \btheta_j; \bmu_j, \brbmu_j),
\]
where
\begin{equation}
\begin{split}
\label{eq:penlogpartiallkd}
\ell_j^{(\rP)}(\bgamma_j, \btheta_j; \bmu_j, \brbmu_j)
&\coloneqq \ell_j(\bgamma_j, \btheta_j) - \sum_{l=1}^p \left\{\brmu^2_{jl} \|\brbD \bgamma_{jl}\|^2_\rF + \mu^2_{jl} \|\bgamma_{jl} \bD^\top\|^2_\rF \right\} \\
% &= \sum_{g=1}^G \ell_{jg}(\bgamma_j, \btheta_j) - \sum_{l=1}^p \left\{\brmu^2_{jl} \|(\brbD\otimes \bI) \vec(\bgamma^\top_{jl})\|^2 + \mu^2_{jl} \|(\brbI\otimes\bD)\vec(\bgamma^\top_{jl}) \|^2 \right\} \\
% &= \sum_{g=1}^G \ell_{jg}(\bgamma_j, \btheta_j) - \sum_{l=1}^p \left\{ \|\brmu_{jl}(\brbD\otimes \bI) \vec(\bgamma^\top_{jl})\|^2 +  \|\mu_{jl}(\brbI\otimes\bD)\vec(\bgamma^\top_{jl}) \|^2 \right\} \\
% &= \sum_{g=1}^G \ell_{jg}(\bgamma_j, \btheta_j) - \|(\brbD \otimes \bI) \bGamma^\top_j \diag(\brbmu_j)\|_\rF^2 - \|(\brbI\otimes\bD) \bGamma^\top_j \diag(\bmu_j)\|_\rF^2 \\
% &= \sum_{g=1}^G \ell_{jg}(\bgamma_j, \btheta_j) - \|\{\brbD\otimes \bI \otimes \diag(\brbmu_j)\} \bgamma_j\|^2 -  \|\{\brbI\otimes\bD \otimes \diag(\bmu_j)\} \bgamma_j\|^2 \\
&= \ell_j(\bgamma_j, \btheta_j) - \bgamma^\top_j \bP_j(\bmu_j, \brbmu_j) \bgamma_j.
\end{split}
\end{equation}
In \eqref{eq:penlogpartiallkd}, $\bmu_j \coloneqq [\mu_{j1}, \mu_{j2}, \ldots{}, \mu_{jp}]^\top$ and $\brbmu_j \coloneqq [\brmu_{j1}, \brmu_{j2}, \ldots{}, \brmu_{jp}]^\top$) denote vectors of smoothing parameters controlling the amount of penalty, $\|\cdot\|_\rF$ denotes the Frobenius norm, and
\begin{multline*}
\bP_j(\bmu_j, \brbmu_j) \coloneqq
\{\brbD \otimes \bI \otimes \diag(\brbmu_j)\}^\top \{\brbD \otimes \bI \otimes \diag(\brbmu_j)\} \\
+ \{\brbI\otimes\bD \otimes \diag(\bmu_j)\}^\top \{\brbI\otimes\bD \otimes \diag(\bmu_j)\},   
\end{multline*}
% $\|\cdot\|$ denotes the $L^2$ norm, 
where $\bI$ (or $\brbI$) is a $K \times K$ (or $\brK \times \brK$) identity matrix, $\diag(\cdot)$ converts a vector into a diagonal matrix, and $\brbD$ (or $\bD$) is a $(\brK-1) \times \brK$ (or $(K-1) \times K$) first-order difference matrix. In what follows, we use tildes to indicate penalized estimates. For example, $\tilde\bgamma_{jl}$ denotes a vector of penalized estimates of $\bgamma_{jl}$.

\subsection{Asymptotics}
\label{s:asymp}
Before presenting the inferential procedures with penalization, we first derive the asymptotic distribution of the penalized estimates $\tilde\beeta_j \coloneqq [\tilde\bgamma^\top_j, \tilde\btheta^\top_j]^\top$. We assume that the knot locations, $K$, $\brK$, $p$, and $q$ remain fixed as the sample size $n \coloneqq \sum_{g=1}^G n_g$ increases. Observe that as $n$ grows, the contribution from the log-partial likelihood in \eqref{eq:penlogpartiallkd} increases. To preserve the degree of smoothness, $\bmu_j$ and $\brbmu_j$ will need to increase at a rate of $O(\sqrt{n})$. Here we consider two cases when the contribution of the penalty term to the penalized score function, $\bP_j(\bmu_j, \brbmu_j) \bgamma_j$, is not necessarily $\mathbf{0}$, but the amount of smoothing (and the introduced bias) shrinks as $n$ increases. First, given two constants $\mu^{(0)}_{jl}$ and $\brmu^{(0)}_{jl}$, if
% $\mu_{jl} = \mu^{(n)}_{jl}$ and $\brmu_{jl} = \brmu^{(n)}_{jl}$ with 
$\mu_{jl}/n^{1/4} \rightarrow \mu^{(0)}_{jl}$ and $\brmu_{jl}/n^{1/4} \rightarrow \brmu^{(0)}_{jl}$ as $n$ increases \citep{gray1992flexible}, then standard derivations imply that $\sqrt{n}(\tilde\beeta_j - \beeta_j)$ is asymptotically normal with a mean estimate
\[
\sqrt{n} \tilde\bb_j \coloneqq \sqrt{n}\left\{\ddot\ell_j^{(\rP)}(\tilde\beeta_j; \bmu_j, \brbmu_j) \right\}^{-1}\bQ_j(\bmu_j, \brbmu_j) \tilde\beeta_j
\]
and a sandwich estimate of variance
\[
n\widetilde\bV_j^{\rS} \coloneqq -n\left\{\ddot\ell_j^{(\rP)}(\tilde\beeta_j; \bmu_j, \brbmu_j) \right\}^{-1}
\ddot\ell_j(\tilde\beeta_j) \left\{\ddot\ell_j^{(\rP)}(\tilde\beeta_j; \bmu_j, \brbmu_j) \right\}^{-1},
\]
where
\[
\ddot\ell_j^{(\rP)}(\beeta_j; \bmu_j, \brbmu_j) 
= \ddot\ell_j(\beeta_j) - \bQ_j(\bmu_j, \brbmu_j)
\]
is the penalized Hessian matrix of \eqref{eq:penlogpartiallkd}, and $\bQ_j(\bmu_j, \brbmu_j)$ is a block diagonal matrix with two blocks $\bP_j(\bmu_j, \brbmu_j)$ and $\mathbf{0}$ (a $q \times q$ matrix). As a second case, if $\mu_{jl}/n^{1/4} \rightarrow 0$ and $\brmu_{jl}/n^{1/4} \rightarrow 0$ as $n$ increases, the variance of $(\tilde\beeta_j - \beeta_j)$ can be well approximated by the inverse of the penalized information matrix, i.e., $\widetilde\bV_j^{\rM} = -\left\{\ddot\ell_j^{(\rP)}(\tilde\beeta_j; \bmu_j, \brbmu_j) \right\}^{-1}$, a model-based variance estimate.

\subsection{Inference}
\label{s:inferpen}
In the presence of penalization, a Wald test statistic associated with the null hypothesis $H^{(t)}_{0}: \bC^{(t)} \vec(\bgamma^\top_{jl}) = \mathbf{0}$ can be written as
\begin{equation}
\label{eq:Waldpdpen}
\{\vec(\tilde\bgamma^\top_{jl})-\tilde\bb_{jl}\}^\top \{\bC^{(t)}\}^\top \left[\bC^{(t)}\bOmega_{jl}\{\bC^{(t)}\}^\top\right]^{-1} \bC^{(t)} \{\vec(\tilde\bgamma^\top_{jl})-\tilde\bb_{jl}\},
\end{equation}
where $\tilde\bb_{jl}$ denotes the $l$th $K\brK$-dimensional subvector of $\tilde\bb_j$, and $\bOmega_{jl}$ denotes an arbitrary $K\brK \times K\brK$ symmetric and positive-definite matrix, e.g., the $l$th diagonal block of $\widetilde\bV_j^\rS$ or $\widetilde\bV_j^\rM$. The distribution of the test statistic \eqref{eq:Waldpdpen} is characterized in \Cref{prop:dist} below. The proof is available in Appendix B.

\begin{proposition}
\label{prop:dist}
Under $H^{(t)}_{0}$, the test statistic \eqref{eq:Waldpdpen} asymptotically follows a distribution characterized by
\[
\sum_{u=1}^{K\brK \times K\brK} \mu_u G^2_u,
\]
where $G_u$'s are independent standard normal random variables, and $\mu_u$'s are the possibly identical eigenvalues of the matrix product of $[\bC^{(t)}\bOmega_{jl}\{\bC^{(t)}\}^\top]^{-1}$ and the variance of $\bC^{(t)} \{\vec(\tilde\bgamma^\top_{jl})-\tilde\bb_{jl}\}$.
\end{proposition}

Similarly as in \Cref{s:inferunpen}, for the null $H^{(\brx)}_{0}: \bC^{(\brx)} \vec(\bgamma^\top_{jl}) = \mathbf{0}$, the corresponding Wald test statistic can be obtained by substituting $\bC^{(t)}$ in \eqref{eq:Waldpdpen} with $\bC^{(\brx)}$. For the null $H^{(\brx)}_{0}: \bC^{(t,\brx)} \vec(\bgamma^\top_{jl}) = \mathbf{0}$, the Wald test statistic can be written by substituting $\bC^{(t)}$ in \eqref{eq:Waldpdpen} with $\bC^{(t,\brx)}$.

\subsection{Cross-validated parameter tuning}
\label{s:CV}
To identify an optimal set of tuning parameters to alleviate model overfitting and the unsmoothness of the estimated effect surface, we consider 5 methods of cross-validation. In the first 4 methods, the entire data sample needs to be partitioned into $F$ subsamples (hereafter folds) of approximately equal sizes. For failure type $j$ and $f = 1, \ldots{}, F$, let $\tilde\beeta_j^{-f}$ be the penalized estimates of $\beeta_j$ based on the complement of fold $f$, and let $\ell_j^f$ and $\ell_j^{-f}$ be the (unpenalized) log-partial likelihood based on fold $f$ and the complement of fold $f$, respectively. A cross-validation error (CVE) for failure type $j$ is then defined in each of the 4 approaches. The last method of generalized cross-validation does not require data partitioning in the calculation of CVE. Optimal tuning parameters can be determined through minimizing the CVE. A comprehensive evaluation of the 5 approaches is presented in \Cref{s:SDApen}.

\subsubsection{Fold-constrained (FC) cross-validated partial likelihood}
In this approach, the CVE is proportional to the sum of fold-specific log-partial likelihood functions in which risk sets are constrained by the corresponding folds, i.e.,
\[
\CVE_j \coloneqq
-2 \sum_{f=1}^F \ell_j^f(\tilde\beeta_j^{-f}).
\]

\subsubsection{Complementary fold-constrained (CFC) cross-validated partial likelihood}
As the name suggests, the CVE is proportional to the sum of complementary fold-constrained log-partial likelihood functions, i.e.,
\[
\CVE_j \coloneqq
-2 \sum_{f=1}^F \{\ell_j(\tilde\beeta_j^{-f}) - \ell_j^{-f}(\tilde\beeta_j^{-f})\}.
\]
This approach was applied in \citet{verweij1993cross} and \citet{simon2011regularization}.
\subsubsection{Unconstrained (UC) cross-validated partial likelihood}
First introduced by \cite{breheny2011coordinate} as cross-validated linear predictors, this approach features risk set construction unconstrained by folds in that fold-specific estimates $\tilde\beeta_j^{-f}$'s are assigned to all units of the sample according to their fold identities. With a slight abuse of notation, the CVE is written as
\[
\CVE_j \coloneqq
-2 \ell_j(\tilde\beeta_j^{-1}, \ldots{}, \tilde\beeta_j^{-F}),
\]
where $\tilde\beeta_j^{-f}$ is assigned to observations of fold $f$.

\subsubsection{Cross-validated deviance residuals (DR)}
\citet{dai2019cross} used the sum of squared deviance residuals \citep{therneau1990martingale} as a criterion of cross-validation in a penalized Cox proportional hazards model. However, their approach cannot be directly applied to a non-proportional hazards model with varying coefficients. To proceed, we first derive the deviance residuals for model \eqref{eq:cause-specifichazard} in the next proposition, the proof of which is available in Appendix C.

\begin{proposition}
\label{prop:DR}
Let $\hat\lambda_{0jg}(\cdot)$ be the estimated baseline hazard function derived from the unpenalized bivariate varying coefficient model. Let 
\[
\tilde M_{jgi} \coloneqq
% \tilde M_{jgi}(\infty, \brX_{gi}) =
\Delta_{jgi} -\exp(\bW_{gi}^\top \tilde\btheta_j^{-f})\int_{0}^{X_{gi}} \exp\left\{\bZ_{gi}^\top \tilde\bbeta_j^{-f}(t, \brX_{gi})\right\} \hat\lambda_{0jg}(t) \,\rd t
\]
be the martingale residual for subject $i$ in the $g$th stratum, where $\tilde\bbeta_j^{-f}(\cdot, \cdot)$ and $\tilde\btheta_j^{-f}$ are the penalized estimates from the corresponding fold $f$ to which subject $i$ in the $g$th stratum belongs. Then the deviance residual for subject $i$ in the $g$th stratum with respect to the $j$th failure type is written as
\[
d_{jgi} \coloneqq \sign(\tilde M_{jgi}) \sqrt{-2\left[\Delta_{jgi} \left\{\bZ_{gi}^\top\tilde\bbeta_j^{-f}(X_{gi}, \brX_{gi}) + \bW_{gi}^\top \tilde\btheta_j^{-f} + \log\int_0^{X_{gi}}  \hat\lambda_{0jg}(t)\, \rd t\right\} + \tilde M_{jgi}\right]}.
\]
\end{proposition}

Given the deviance residuals in \Cref{prop:DR}, the CVE can be written as
\[
\CVE_j \coloneqq
\sum_{g=1}^G \sum_{i=1}^{n_g} d^2_{jgi}.
\]

\subsubsection{Generalized cross-validation (GCV)}
Extending the approach of \citet{yan2012model} to this setting with bivariate varying coefficients, we can write the CVE for the $j$th failure type as
\[
\mathrm{CVE}_j
= - \frac{\ell_j(\beeta_j)}{n(1-f_j(\bmu_j, \brbmu_j)/n)^2},
\]
where $f_j(\bmu_j, \brbmu_j) \coloneqq \tr\left(\{\ddot\ell_j^{(\rP)}(\beeta_j; \bmu_j, \brbmu_j)\}^{-1}\ddot\ell_j(\beeta_j)\right)$, i.e., the number of effective parameters \citep{yan2012model}, or the ``degrees of freedom'' of the model \citep{gray1992flexible}.

\section{Simulation experiments}
\label{s:SDA}

\subsection{Unpenalized approach}
Following the approach in \Cref{s:estandinfunpen}, we assessed the bivariate varying coefficient model for competing risks via simulation experiments. Since distinct types of competing risks can be analyzed separately within a cause-specific hazard framework, we focused on a single event type and dropped the subscript $j$ to allow simplified notation. Therefore, no stratification was used in the data generating process.

In each simulation scenario, a number (100 or 1,000) of independent data replicates were generated with the sample size varying from 1,000 to 10,000. For each sample unit, two covariates (corresponding to $\bZ_{gi}$ and $\bW_{gi}$ in \Cref{s:model}) were drawn from a bivariate normal distribution with zero mean, one variance, and correlation $\rho = 0.6$. Two coefficients were set as $\beta_1(t, \brx) = \sin(3\pi t/4)\exp(-0.5\brx)$ and $\beta_2 = 1$, with event time $t$ varying from 0 to 30, and calendar time $\brx$ varying from 0 to 50. Underlying event times were determined via a root-finding procedure based on the cause-specific hazard function in \Cref{s:model} \citep{beyersmann2009simulating}. Calendar times were drawn from a uniform distribution bounded by 0 and 50. Censoring times were sampled from a uniform distribution bounded by 0 and 30. Observed event times were determined as the minimum of the underlying event and censoring time pairs.

\begin{figure}
    \centering
    \begin{subfigure}[t]{0.49\textwidth}
    \centering
    \includegraphics[width=\linewidth]{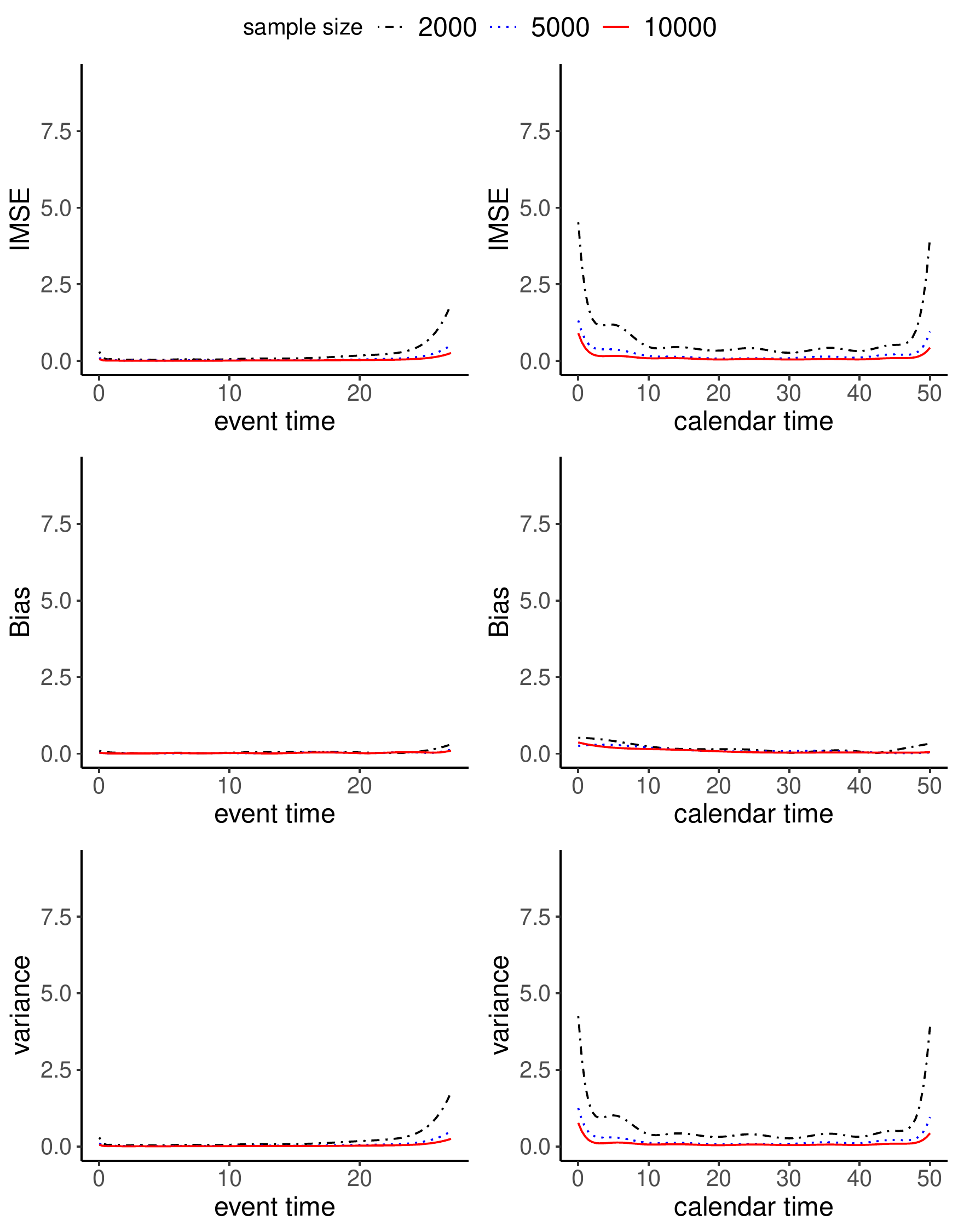} 
    \caption{IMSE, bias, and variance} \label{fig:unpen_imse_pd_cal}
    \end{subfigure}
    \begin{subfigure}[t]{0.49\textwidth}
    \centering
    \includegraphics[width=\linewidth]{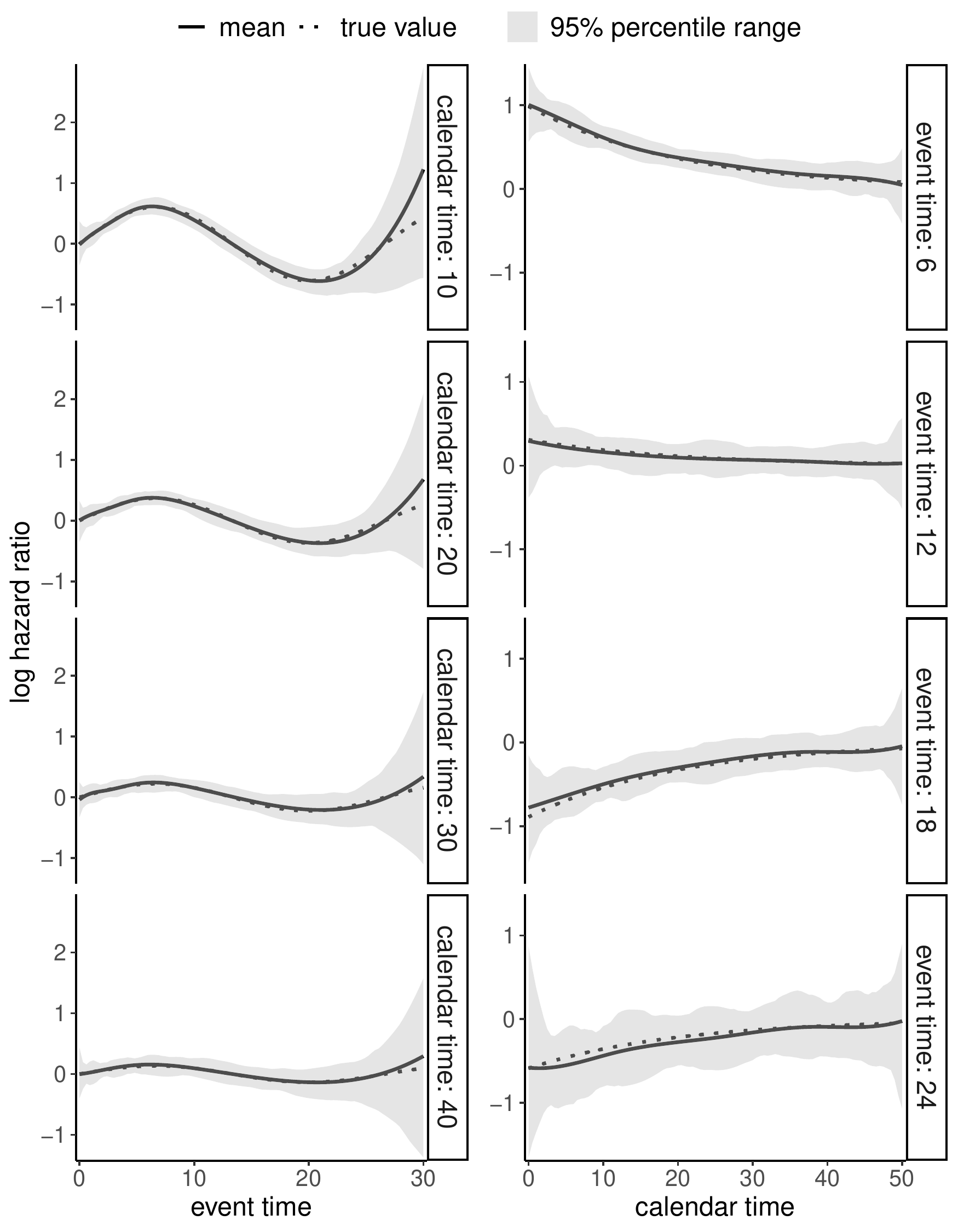} 
    \caption{Mean versus true value}
    \label{fig:unpen_mean_vs_true_pd_cal}
    \end{subfigure}
    \label{fig:unpen_acc}
    \caption{(a) Integrated mean squared error (IMSE), average bias, and average variance of the estimated surface $\hat\beta_1(t, \brx)$ with varied sample sizes on event and calendar timescales. In each scenario, 100 data replicates were generated. On both timescales, $K = \brK = 7$ cubic ($d = \brd = 3$) B-spline functions form a basis. True values are $\beta_1(t, \brx) = \sin(3\pi t/4)\exp(-0.5\brx)$ and $\beta_2 = 1$. (b) Mean and 95\% percentile range (2.5th and 97.5th percentiles as lower and upper limits) of pointwise estimates of $\beta_1(t, \brx)$ at selected event times and calendar times. In each scenario, 100 data replicates were generated with sample size equal to 10,000. On both timescales, $K = \brK = 7$ cubic ($d = \brd = 3$) B-spline functions form a basis. True values are $\beta_1(t, \brx) = \sin(3\pi t/4)\exp(-0.5\brx)$ and $\beta_2 = 1$. An unpenalized approach was used in (a) and (b).}
\end{figure}

\Cref{fig:unpen_imse_pd_cal} presents the integrated mean squared error (IMSE), bias, and variance (all averaged over a grid of 100 evenly spaced points across either event or calendar time) with respect to the bivariate varying coefficient $\beta_1(t, \brx)$ with the sample size growing from 2,000 to 10,000. The three metrics were calculated based on 100 data replicates. The coefficient $\beta_2$ was treated as a time-invariant parameter in model fitting. On the event timescale, the IMSE becomes higher as event time increases, due to the fact that the shrinking risk set leads to fewer remaining units in the sample and hence less accurate estimation. As the sample size grows, the IMSE curve shifts downward and the IMSE is substantially reduced towards the end of follow-up. On the calendar timescale, the IMSE is higher on both ends and the curve becomes lower as the sample size increases from 2,000 to 10,000. Moreover, a comparison between the second and third row of \Cref{fig:unpen_imse_pd_cal} suggests that the IMSE on both timescales is predominantly determined by the variance component.  As a complement to \Cref{fig:unpen_imse_pd_cal}, \Cref{fig:unpen_mean_vs_true_pd_cal} provides additional evidence on estimation, with the sample size fixed at 10,000. Throughout all panels of distinct event and calendar times, the mean estimated curve tracks closely with the true effect curve, demonstrating the accurate estimation of the extended proximal Newton algorithm.

\begin{figure}
    \centering
    \begin{subfigure}[t]{0.7\textwidth}
    \centering
    \includegraphics[width=\linewidth]{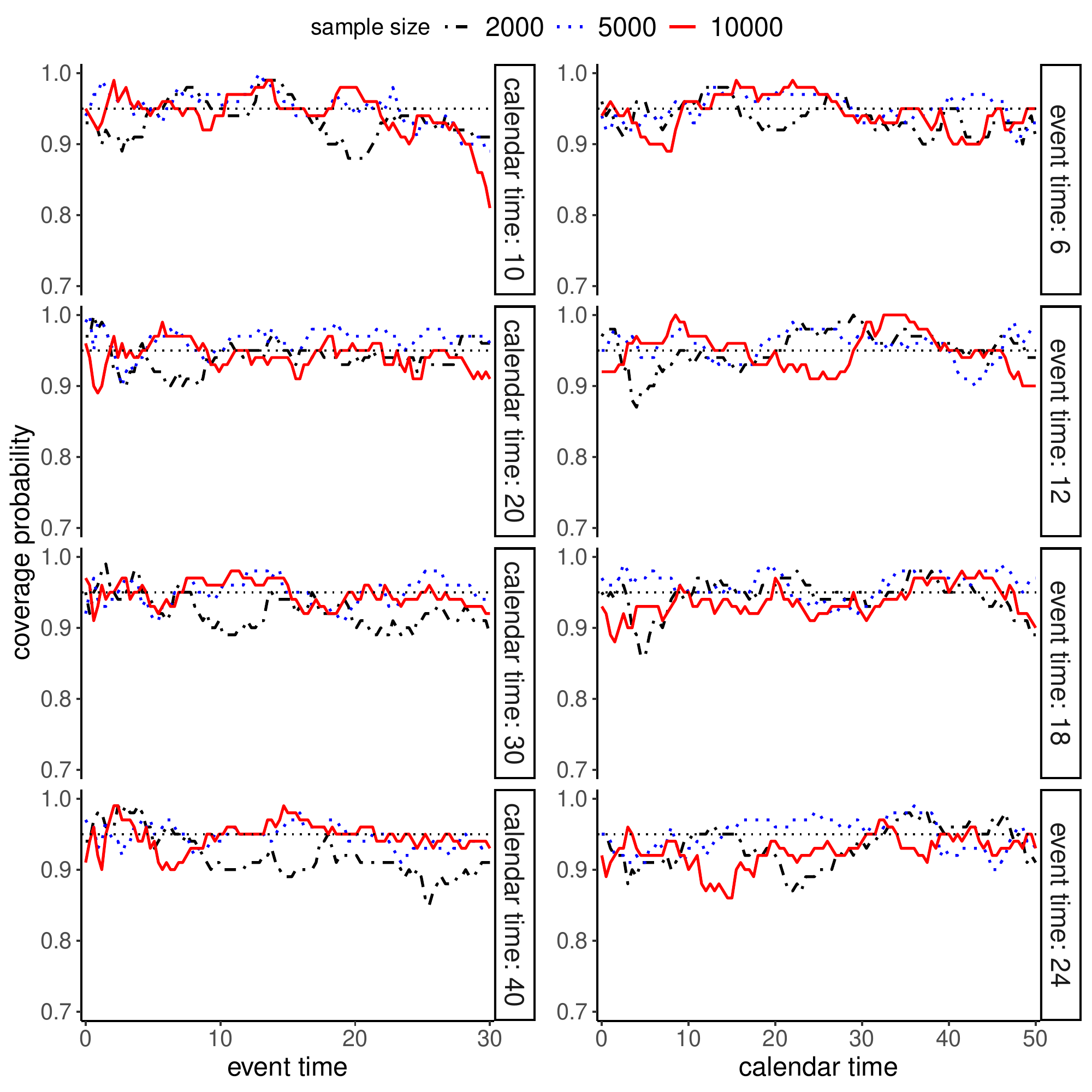} 
    \caption{Coverage probability} \label{fig:unpen_CP_pd_cal}
    \end{subfigure}
    \begin{subfigure}[t]{0.67\textwidth}
    \centering
    \includegraphics[width=\linewidth]{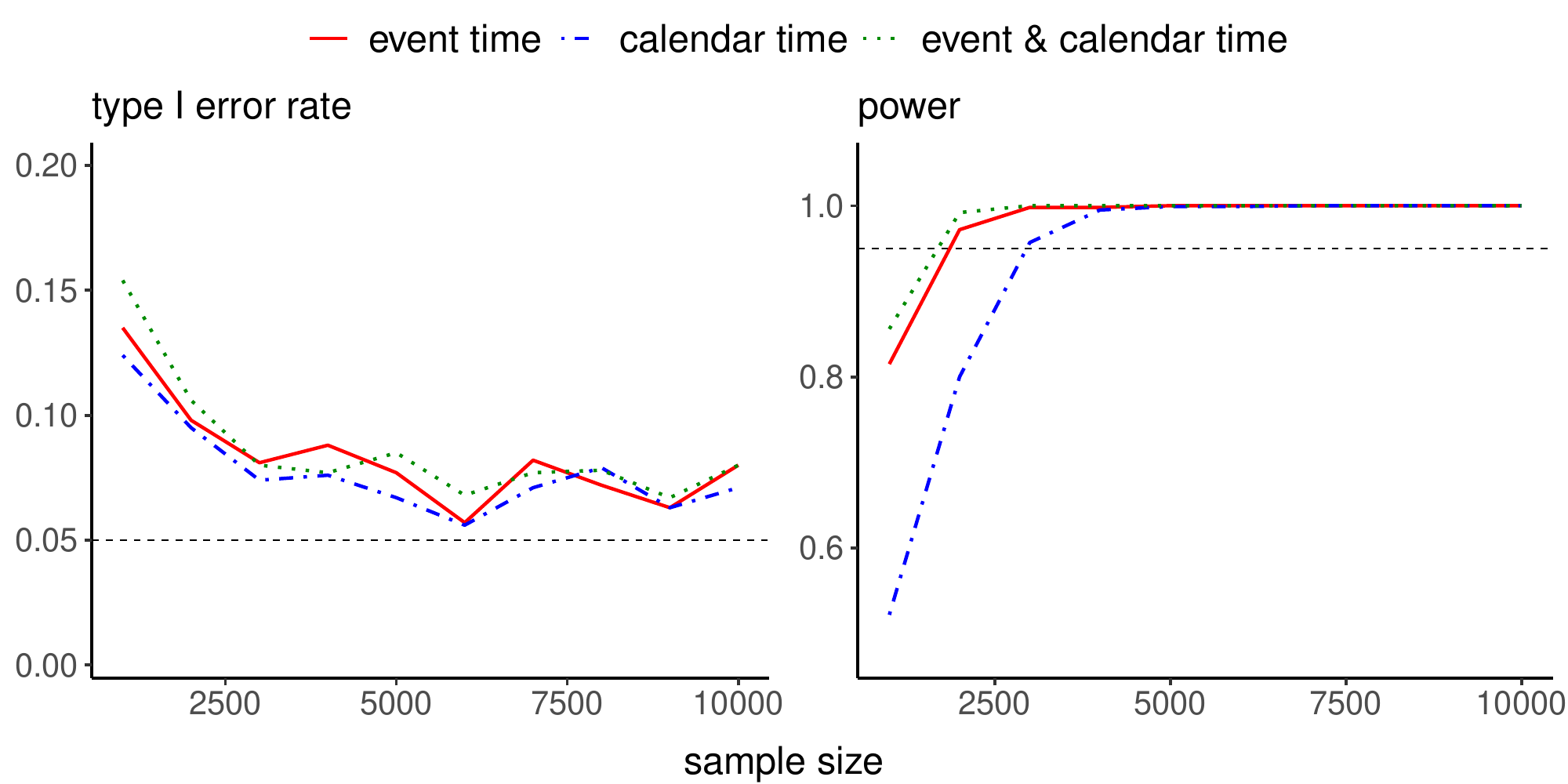} 
    \caption{Type I error rate and power} \label{fig:unpen_errpower_pd_cal}
    \end{subfigure}
    \caption{(a) Coverage probability curves of $\beta_1(t, \brx)$ via pointwise 95\% confidence intervals on event and calendar time scales, with varied sample sizes. In each scenario, 100 data replicates were generated with sample size equal to 10,000. On both timescales, $K = \brK = 7$ cubic ($d = \brd = 3$) B-spline functions form a basis. True values are $\beta_1(t, \brx) = \sin(3\pi t/4)\exp(-0.5\brx)$ and $\beta_2 = 1$. (b) Type I error rate and power curves for tests of univariate and bivariate variation with varied sample sizes. In each scenario, 1,000 data replicates were generated. On both timescales, $K = \brK = 7$ cubic ($d = \brd = 3$) B-spline functions form a basis. True values are $\beta_1(t, \brx) = 1$ and $\beta_2 = 1$ in the left panel, and $\beta_1(t, \brx) = \sin(3\pi t/4)\exp(-0.5\brx)$ and $\beta_2 = 1$ in the right panel. An unpenalized approach was used in (a) and (b).}
\end{figure}

At different event and calendar times, we compared coverage probability (CP) curves of $\beta_1(t, \brx) = \sin(3\pi t/4)\exp(-0.5\brx)$ with varied sample sizes in \Cref{fig:unpen_CP_pd_cal}. Pointwise 95\% confidence intervals were used throughout all panels. Overall, all curves remained around the 0.95 reference line, except that the CP dropped below 0.8 toward the end of the event time period with calendar time equal to 10 and sample size equal to 10,000. In \Cref{fig:unpen_errpower_pd_cal}, we evaluated three tests of univariate and bivariate variation with respect to $\beta_1(t, \brx)$, where the sample size varied from 2,000 to 10,000. As expected \citep{wu2022scalable}, curves of type I error rate were sloping downward with the sample size, while the rates remained slightly higher than 0.05 as the sample size exceeded 5,000. The power grew dramatically until the sample size reached 4,000, and then remained higher than 0.95 afterward.

\subsection{Penalized approach}
\label{s:SDApen}

\begin{figure}
    \centering
    \begin{subfigure}[t]{0.49\textwidth}
    \centering
\includegraphics[width=\linewidth]{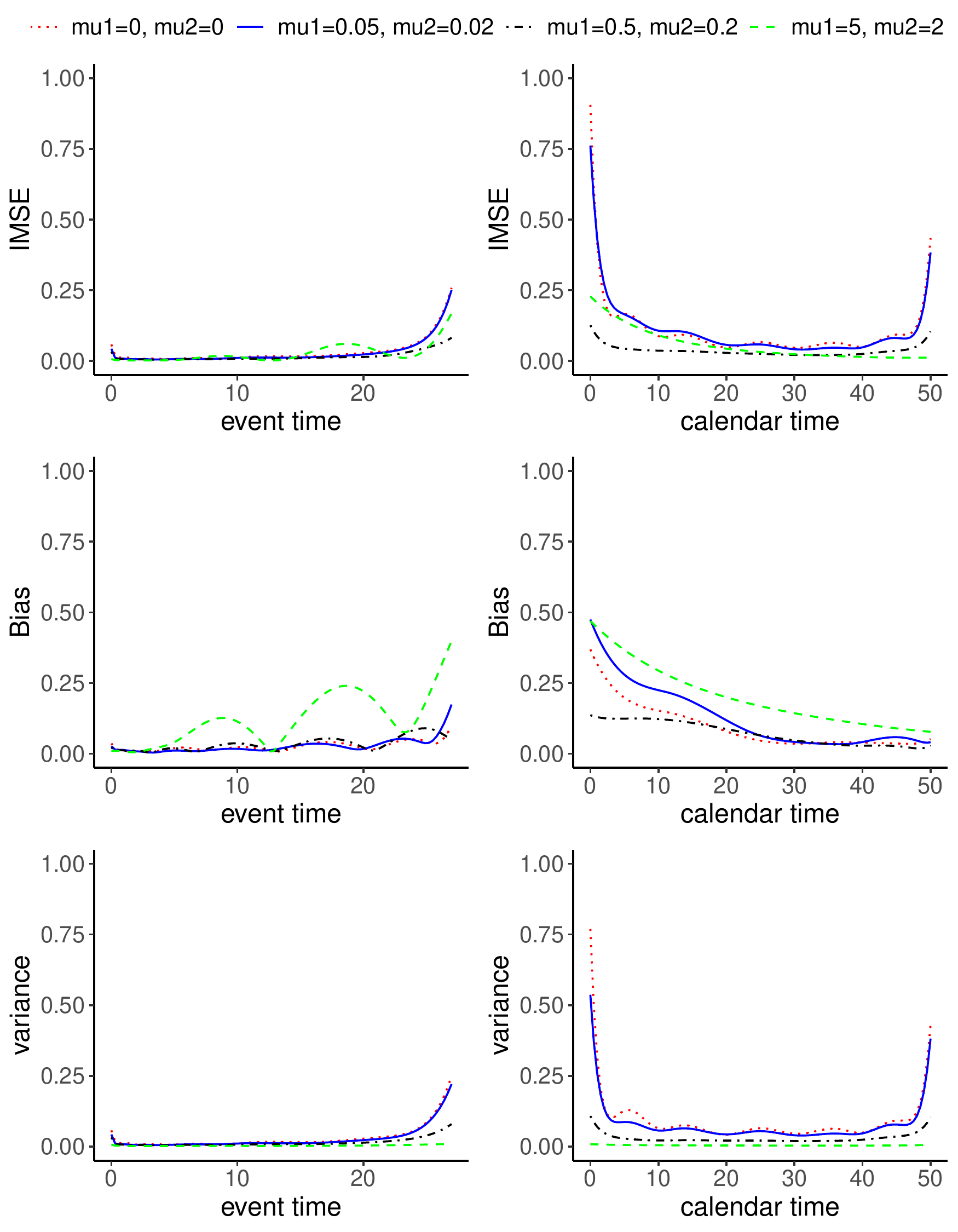} 
    \caption{IMSE, bias, and variance} \label{fig:pen_imse_pd_cal}
    \end{subfigure}
    \begin{subfigure}[t]{0.49\textwidth}
    \centering
    \includegraphics[width=\linewidth]{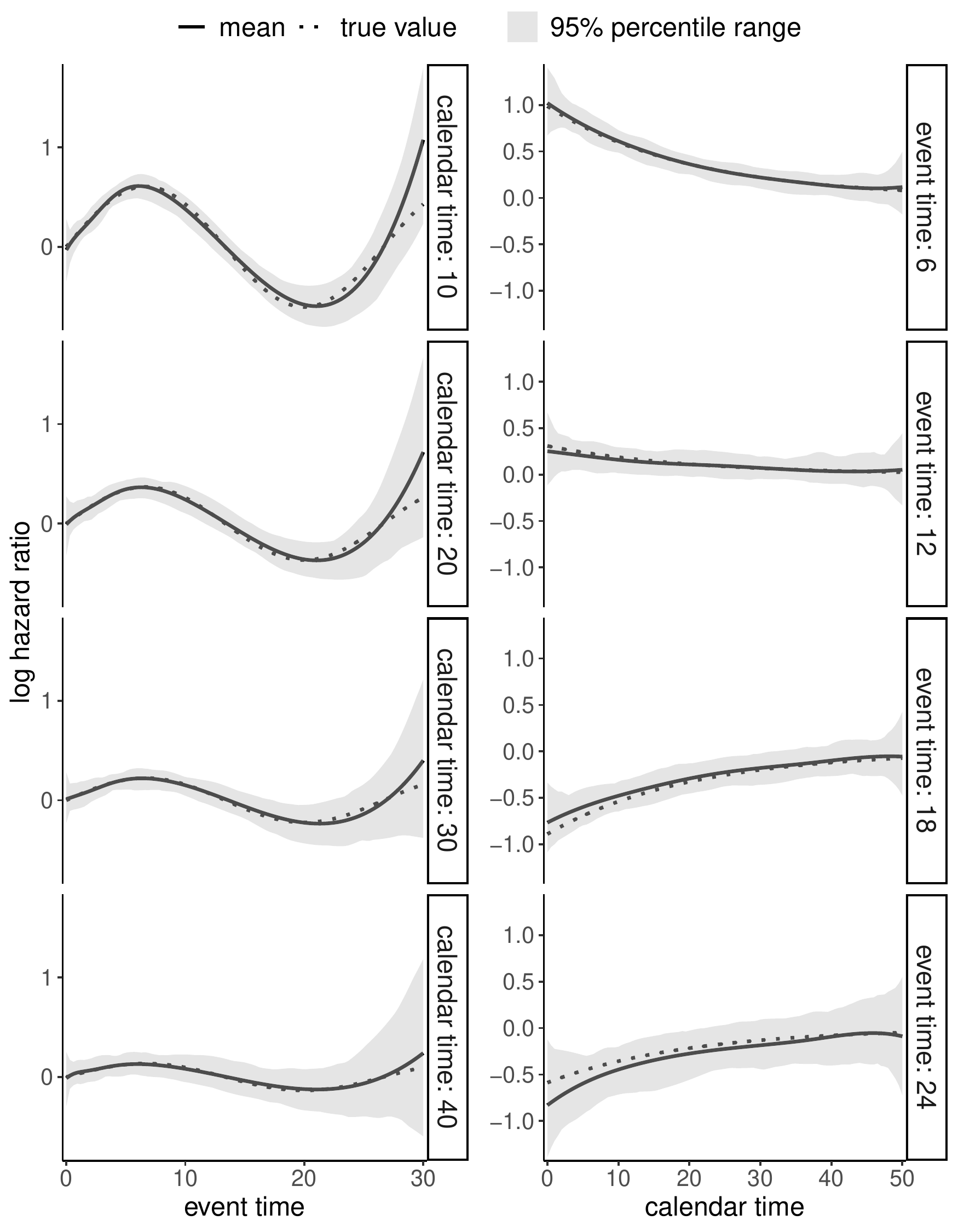}
    \caption{Mean versus true value}
    \label{fig:pen_mean_vs_true_pd_cal}
    \end{subfigure}
    \caption{(a) Integrated mean squared error (IMSE), average bias, and average variance of the estimated surface $\hat\beta_1(t, \brx)$ with sample size fixed at 10,000. In each scenario, 100 data replicates were generated. On both timescales, $K = \brK = 7$ cubic ($d = \brd = 3$) B-spline functions form a basis. True values are $\beta_1(t, \brx) = \sin(3\pi t/4)\exp(-0.5\brx)$ and $\beta_2 = 1$. Various levels of penalization were introduced to $\beta_1(\cdot, \cdot)$, where \textsf{mu1} and \textsf{mu2} denote tuning parameters for calendar and event time, respectively, as in \eqref{eq:penlogpartiallkd}. (b) Mean and 95\% percentile range (2.5th and 97.5th percentiles as lower and upper limits) of pointwise estimates of $\beta_1(t, \brx)$ at selected event times and calendar times. In each scenario, 100 data replicates were generated with sample size equal to 10,000. On both timescales, $K = \brK = 7$ cubic ($d = \brd = 3$) B-spline functions form a basis. True values are $\beta_1(t, \brx) = \sin(3\pi t/4)\exp(-0.5\brx)$ and $\beta_2 = 1$. Only the optimal case in Part (a), i.e., \textsf{mu1}=0.5 and \textsf{mu2}=0.2, was considered.\label{fig:pen_acc}}
\end{figure}

In similar simulation settings, we evaluated the difference-based anisotropic penalization and corresponding tests of effect variation. With sample size fixed at $n=10\text{,}000$, \Cref{fig:pen_imse_pd_cal} shows the IMSE, bias, and variance (again, averaged over a grid a 100 evenly spaced points on either time scale) of the estimated effect surface $\hat\beta_1(t, \brx)$ on two timescales with different pairs of tuning parameters $\mu$ and $\brmu$, where the unpenalized approach with $\mu = 0$ and $\brmu = 0$ is included as a reference. Across both event and calendar time, the IMSE was the highest when the penalty was minimal ($\mu = 0.02$ and $\brmu = 0.05$). As the penalty became more prominent ($\mu = 0.2$ and $\brmu = 0.5$), the IMSE decreased at first, especially on the calendar timescale. When $\mu = 2$ and $\brmu = 5$, the IMSE rebounded substantially. As for bias, a higher penalty level led to a higher bias across event time, while the bias remained lowest with $\mu = 0.2$ and $\brmu = 0.5$. Unsurprisingly, a higher level penalty was associated with lower variance for both timescales. This result suggests that $\mu = 0.2$ and $\brmu = 0.5$ are the optimal pair among the four. This pair was applied exclusively in \Cref{fig:pen_mean_vs_true_pd_cal}, where the curves of true values were compared to the curves of mean estimates. In all panels, the two curves tracked closely, except toward the end of the event time.

\begin{figure}
    \centering
    \begin{subfigure}[t]{0.85\textwidth}
    \centering
    \includegraphics[width=\linewidth]{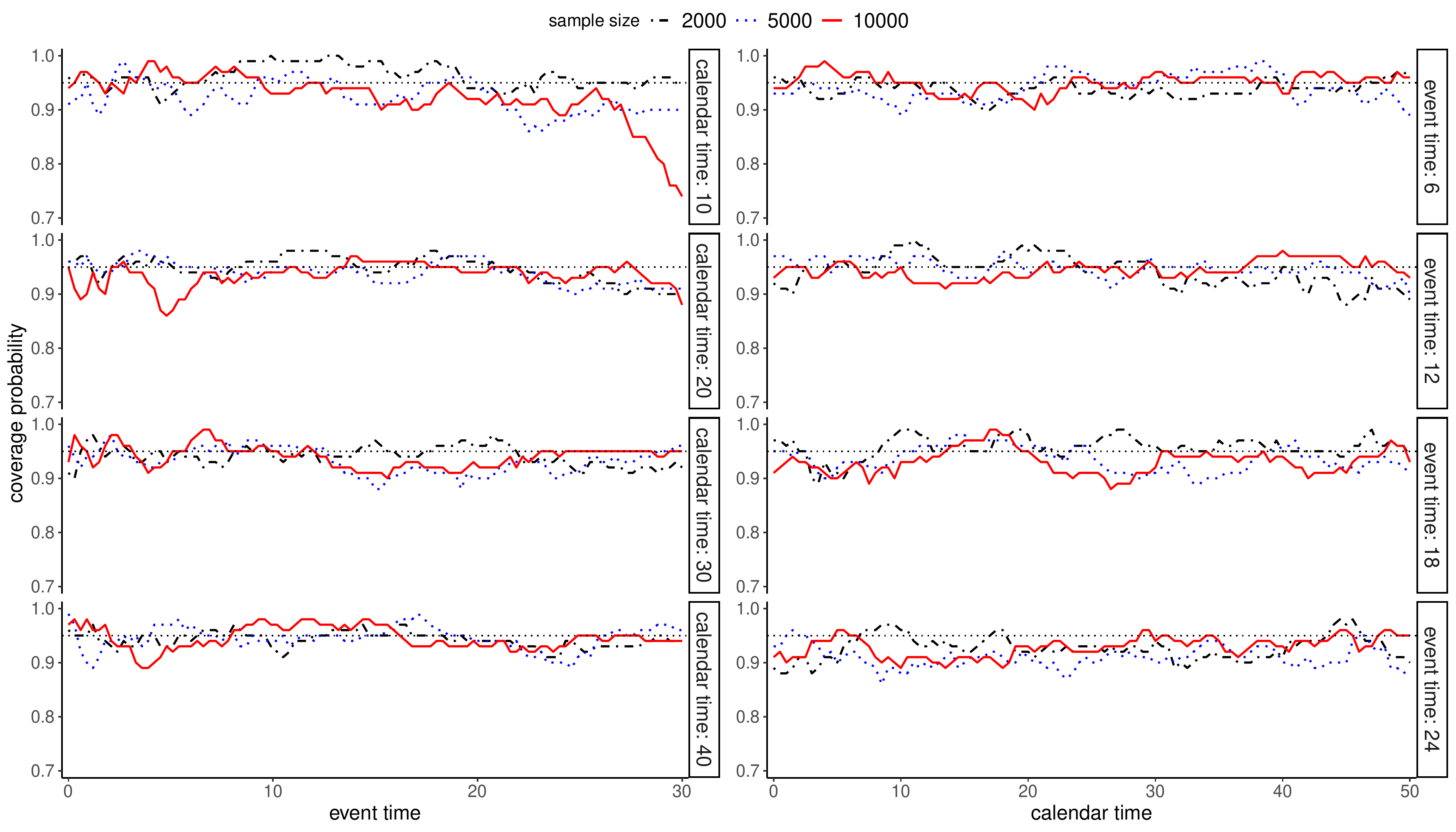} 
    \caption{Coverage probability} \label{fig:pen_CP_pd_cal}
    \end{subfigure}
    \begin{subfigure}[t]{0.85\textwidth}
    \centering
    \includegraphics[width=\linewidth]{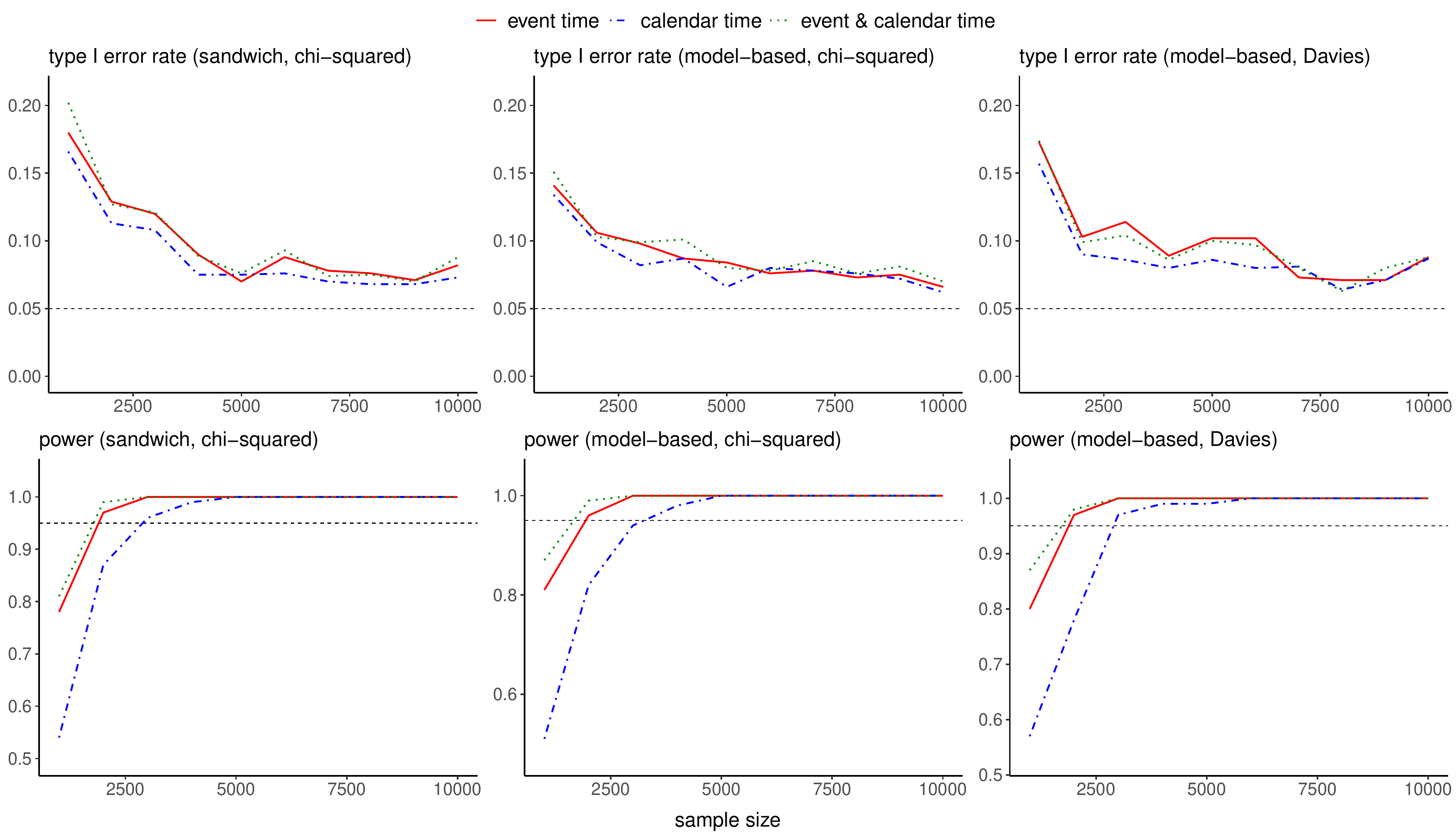} 
    \caption{Type I error rate and power} \label{fig:pen_errpower_pd_cal}
    \end{subfigure}
    \caption{(a) Coverage probability curves of $\beta_1(t, \brx)$ via pointwise 95\% confidence intervals at varied event time, calendar time, and sample sizes. In each scenario, 100 data replicates were generated with sample size $n = 10,000$. True values are $\beta_1(t, \brx) = \sin(3\pi t/4)\exp(-0.5\brx)$ and $\beta_2 = 1$. (b) Type I error rate and power curves for tests of univariate and bivariate variation with different test statistics and varied sample sizes. In each scenario, 1,000 data replicates were generated. True values are $\beta_1(t, \brx) = 1$ and $\beta_2 = 1$ in the top 3 panels, and $\beta_1(t, \brx) = \sin(3\pi t/4)\exp(-0.5\brx)$ and $\beta_2 = 1$ in the bottom 3 panels. In the first and second column, a sandwich and a model-based variance estimator were used with test statistics approximately following a chi-squared distribution. In the third column, the test statistic in \citet{gray1992flexible} was compared with a distribution of a linear combination of chi-squared random variables \citep{davies1980algorithm}. In Parts (a) and (b), 7 cubic B-splines form a basis on both timescales, and tuning parameters vary with sample size, i.e., $\mu = n^{1/8}/500$ and $\brmu = n^{1/8}/200$. \label{fig:pen_CPerrpower}}
\end{figure}

\Cref{fig:pen_CPerrpower} presents the CP, type I error rate, and power with varying sample sizes. To allow tuning parameters to vary with sample size, we set $\mu = 0.002n^{1/8}$ and $\brmu = 0.005n^{1/8}$, corresponding to the second case in \Cref{s:asymp}. Across event and calendar time, the CP curve fluctuated closely around the 0.95 reference line, except that the CP dropped to 0.75 toward the end of the event time period with calendar time equal to 10 and sample size equal 10,000 (\Cref{fig:pen_CP_pd_cal}, top left panel). In each column of \Cref{fig:pen_errpower_pd_cal}, we adopted a distinct construction of the test statistics based on \eqref{eq:Waldpdpen}, and considered three tests of variation, jointly and separately. In the first and second columns, the sandwich and model-based variance estimators, respectively, were employed to determine $\bOmega$ and the variance of $\vec(\tilde\bgamma^\top)-\tilde\bb$, respectively, so that the test statistics approximately followed the chi-squared distribution. In the third column, the model-based estimate was used to form $\bOmega$, while the variance of $\vec(\tilde\bgamma^\top)-\tilde\bb$ was estimated via the sandwich estimator. The resulting test statistics, similar to the one in \citet{gray1992flexible}, approximately followed a distribution characterized by a linear combination of chi-squared random variables \citep{davies1980algorithm}. This distribution was implemented via the package \textsf{CompQuadForm} \citep{ldm2017compquadform}. We observed that the third construction generally led to higher type I error rates than the other two. When sample size was up to 3,000, the model-based test statistics gave lower type I error rates; when sample size exceeded 3,000, the sandwich test statistics overall resulted in slightly lower type I error rates. All three constructions were associated with sufficiently high power with sample size greater than or equal to 3,000.

To compare the five methods of cross-validation via simulations in \Cref{s:CV}, we generated 100 pairs of training and testing data replicates for each sample size $n$ (varying from 2,000 to 5,000). A 5-by-5 grid of tuning parameters was formed such that $\mu/\sqrt{n}$ and $\brmu/\sqrt{n}$ varied from $10^{-5}$ to $10^{-1}$. All five methods were applied to a training copy to obtain an optimal pair of tuning parameters and penalized estimates. The training data were split into four folds whenever data partitioning was necessary. The penalized estimates were then applied to both training and testing replicates in the calculation of $-2\ell$ ($\ell$ denoting the unpenalized log partial likelihood) and the average IMSE, two measures of predictive accuracy used for evaluating the five methods. The distribution of selected tuning parameters is reported in the Appendix D, and $-2\ell$ and average IMSE for all methods are tabulated in \Cref{tab:CV}. The method of cross-validated deviance residuals (DR) led to the lowest $-2\ell$ when the sample size of the training data was less than 5,000, or when the sample size of the testing data was 3,000 or 5,000; it also led to the lowest average IMSE when the sample size of the training data was 4,000 or 5,000. In contrast, the generalized cross-validation was associated with the highest $-2\ell$ for both training and testing data across different sample sizes; it also gave the highest average IMSE except when the sample size was 2,000. Although DR overall achieved the highest predictive accuracy, its advantage over the other 3 data-partitioning cross-validation methods was not significant.

\begin{table}[htbp]
\caption{A simulation-based comparison of five cross-validation methods: fold-constrained (FC), complementary fold-constrained (CFC), and fold-unconstrained (UC) cross-validated partial likelihood, cross-validated deviance residuals (DR), and generalized cross-validation (GCV). In each scenario, 100 training and validation data replicates were generated independently. Each cross-validation method was applied to the training data replicate to obtain the penalized estimates. The estimates were then applied to the training and validation data separately to calculate $-2\ell$ (Panel A), where $\ell$ denotes the unpenalized log partial likelihood, and to the training data to calculate average integrated mean squared error (IMSE, Panel B). For IMSE, the average was taken across 10,201 different combinations of event and calendar time. True values were $\beta_1(t, \brx) = \sin(3\pi t/4)\exp(-0.5\brx)$ and $\beta_2 = 1$. Standard deviations are provided in parentheses. \label{tab:CV}}
\centering
%\footnotesize
\resizebox{\columnwidth}{!}{%
\begin{tabular}{@{}cccccccccccc@{}}
\toprule
\toprule
\multicolumn{12}{c}{Panel A: $-2\ell$} \\
\midrule[0.4pt]
\makecell{\multirow{2}{*}{sample size}} & \multicolumn{5}{c}{training} & &
\multicolumn{5}{c}{testing} \\
\cmidrule[0.4pt]{2-6} \cmidrule[0.4pt]{8-12}
& FC & CFC & UC & DR & GCV & &
FC & CFC & UC & DR & GCV \\
\midrule[0.4pt]
\multirow{2}{*}{2000} & 12951.11 & 12950.53 & 12951.02 & 12949.77 & 12978.45 & & 11917.90 & 11917.24 & 11917.69 & 11917.68 & 11927.87 \\
 & (2636.17) & (2634.96) & (2634.98) & (2635.09) & (2635.43) & & (11.80) & (10.72) & (10.90) & (10.85) & (4.70) \\
\multirow{2}{*}{3000} & 20729.73 & 20729.64 & 20729.66 & 20729.58 & 20769.08 & & 25852.29 & 25852.16 & 25852.21 & 25852.12 & 25871.91 \\
& (4174.25) & (4174.46) & (4174.56) & (4174.55) & (4176.98) & & (16.02) & (16.01) & (15.99) & (15.98) & (8.77) \\
\multirow{2}{*}{4000} & 28696.21 & 28696.33 & 28696.21 & 28695.57 & 28746.12 & & 26794.12 & 26794.10 & 26794.11 & 26794.34 & 26831.69 \\
& (5753.25) & (5752.60) & (5752.59) & (5753.07) & (5757.26) & & (10.51) & (10.49) & (10.50) & (10.57) & (10.30) \\
\multirow{2}{*}{5000} & 37036.32 & 37035.18 & 37035.27 & 37036.26 & 37095.90 & & 54238.77 & 54239.27 & 54239.44 & 54238.45 & 54279.14 \\
& (7439.08) & (7439.23) & (7439.38) & (7439.86) & (7451.81) & & (26.10) & (26.37) & (26.19) & (26.85) & (14.08) \\
\midrule
\multicolumn{12}{c}{Panel B: average IMSE} \\
\cmidrule[0.4pt]{4-9}
 & & & \makecell{\multirow{2}{*}{sample size}} & \multicolumn{5}{c}{training} & & & \\
\cmidrule[0.4pt]{5-9}
& & & & FC & CFC & UC & DR & GCV & & & \\
\cmidrule[0.4pt]{4-9}
 & & & \multirow{2}{*}{2000} & 0.1023 & 0.1019 & 0.1041 & 0.1104 & 0.0812 & & & \\
 & & & & (0.1193) & (0.1137) & (0.1148) & (0.1170) & (0.0124) & & & \\
 & & & \multirow{2}{*}{3000} & 0.0697 & 0.0710 & 0.0706 & 0.0726 & 0.0775 & & &\\
 & & & & (0.0679) & (0.0684) & (0.0685) & (0.0678) & (0.0100) & & &\\
 & & & \multirow{2}{*}{4000} & 0.0714 & 0.0703 & 0.0680 & 0.0674 & 0.0747 & & & \\
 & & & & (0.0917) & (0.0917) & (0.0904) & (0.0917) & (0.0098)  & & & \\
 & & & \multirow{2}{*}{5000} & 0.0562 & 0.0567 & 0.0570 & 0.0550 & 0.0729 & & & \\
 & & & & (0.1193) & (0.1137) & (0.1148) & (0.1170) & (0.0124) & & & \\
\bottomrule
\end{tabular}
}
\end{table}

\section{Applications to dialysis patients amidst COVID-19}
\label{s:RDA}
To better understand the dynamics of the COVID-19 effect on dialysis patients, we applied the bivariate varying coefficient model to two large-scale retrospective studies, both having data abstracted from the CMS clinical and administrative database (primarily based on the Renal Management Information System, CROWNWeb facility-reported clinical and administrative data, the Medicare Enrollment Database, and Medicare claims data). In both studies, the interest was in the impact of an in-hospital COVID-19 diagnosis on the outcomes of dialysis patients. Information on in-hospital COVID-19 diagnosis was mainly obtained from Medicare inpatient and physician/supplier claims. An in-hospital COVID-19 diagnosis was confirmed if the patient's inpatient or physician/supplier claim associated with the hospitalization had either of the two diagnosis codes of the International Classification of Diseases, 10th Revision: B97.29 or U07.1 \citep{wu2021covid19}. In addition to COVID-19, a comprehensive list of patient demographics, clinical characteristics, and prevalent comorbidities were considered as baseline risk factors.

\subsection{Postdischarge outcomes}
\label{s:postdischarge}
In the first study, outcomes of primary interest were all-cause unplanned acute-care-hospital readmission and death within 30 days of hospital discharge. This study consisted of 436,745 live acute-care hospital discharges of 222,154 Medicare beneficiaries on dialysis from 7,871 Medicare-certified dialysis facilities between January 1, 2020 and October 31, 2020. Discharges from non-acute care hospitals, discharges with in-hospital death, and discharges with discharge-day outcomes were excluded from the data, along with other administrative exclusions.

\begin{figure}
    \centering
    \begin{subfigure}[t]{0.7\textwidth}
        \centering
        \includegraphics[width=\linewidth]{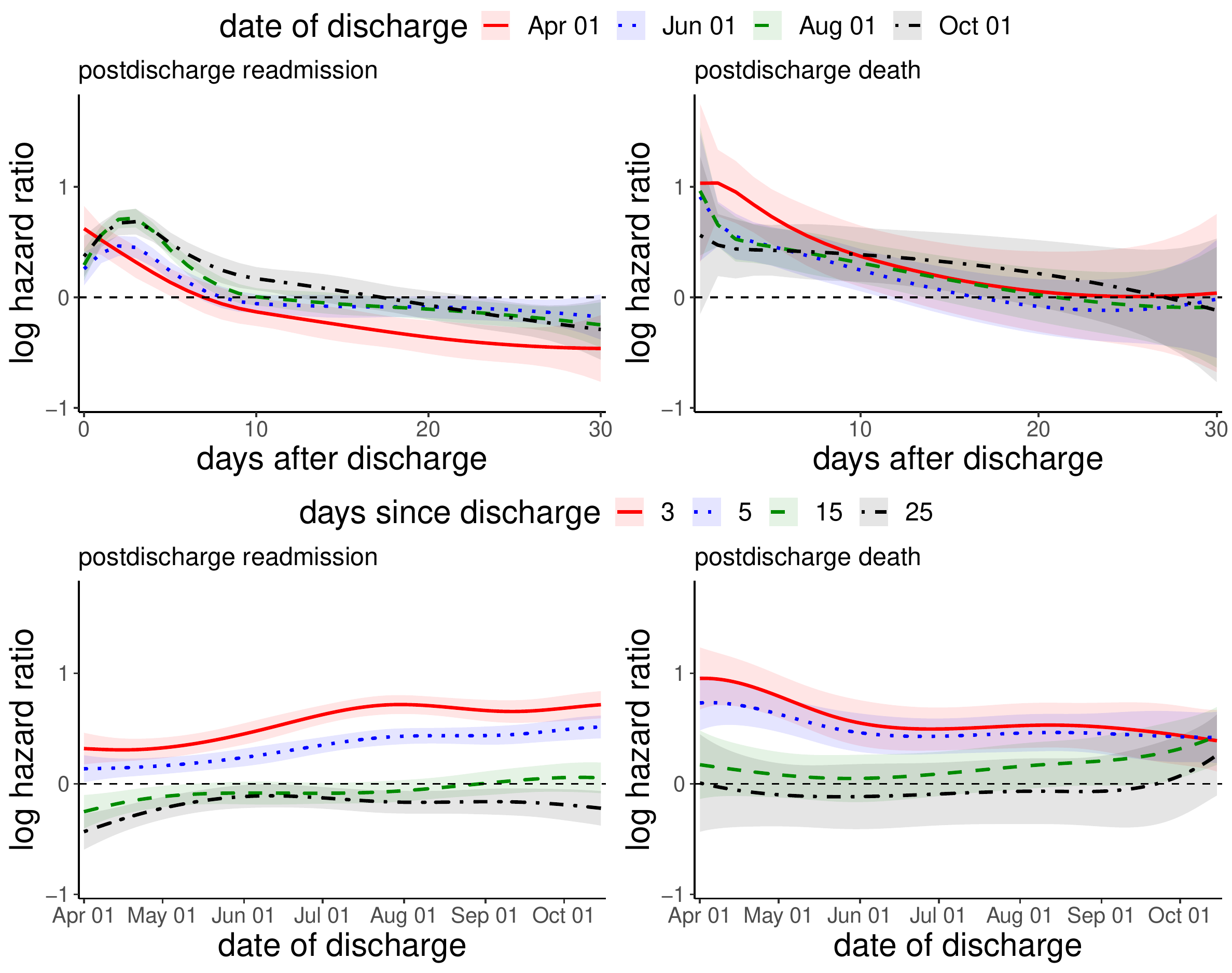} 
    \label{fig:postdischargepdcal}
    \end{subfigure}
    
    \begin{subfigure}[t]{0.35\textwidth}
        \centering
        \includegraphics[width=\linewidth]{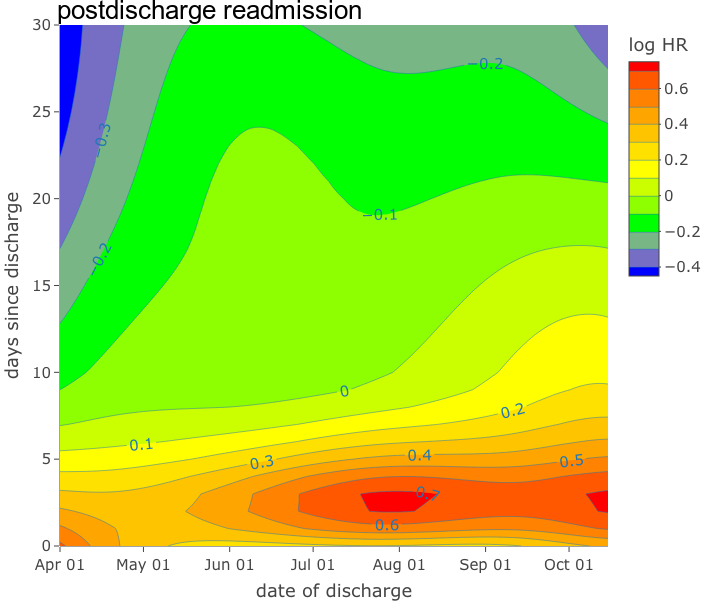} 
        \label{fig:postdischargeUHRcontour}
    \end{subfigure}
   \begin{subfigure}[t]{0.35\textwidth}
    \centering
        \includegraphics[width=\linewidth]{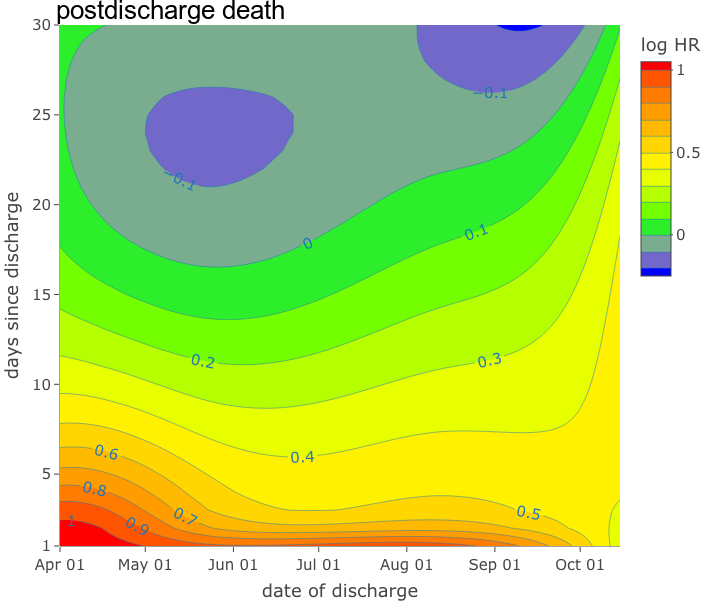}
        \label{fig:postdischargedeathcontour}
    \end{subfigure}
    
    \begin{subfigure}[t]{0.35\textwidth}
    \centering
        \includegraphics[width=\linewidth]{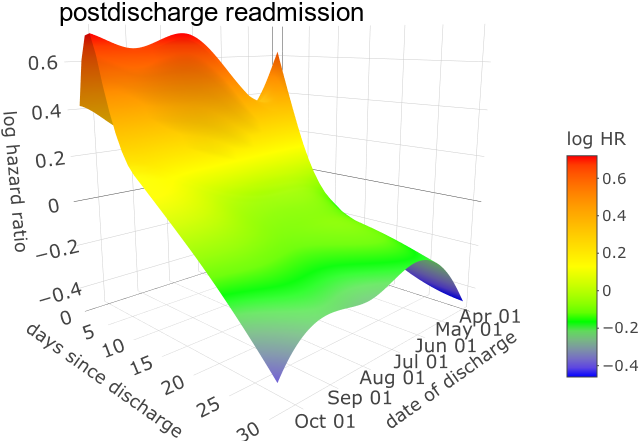} 
        \label{fig:postdischargeUHR3d}
    \end{subfigure}
    \begin{subfigure}[t]{0.35\textwidth}
        \centering
        \includegraphics[width=\linewidth]{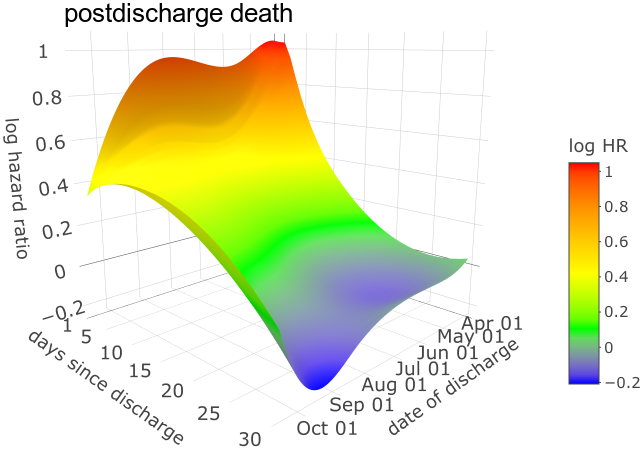} 
        \label{fig:postdischargedeath3d}
    \end{subfigure}
    \caption{Bivariate variation of log hazard ratios with respect to in-hospital COVID-19 diagnosis for 30-day postdischarge readmission and death. Included in the sample were 436,745 live hospital discharges of 222,154 Medicare beneficiaries on dialysis from 7,871 Medicare-certified dialysis facilities from January 1, 2020 to October 31, 2020. Ribbons in the top four panels indicate 95\% confidence intervals. Panels in the third and fourth rows are contour and surface plots, respectively. \label{fig:postdischargeRDA}}
\end{figure}

The 8 panels of \Cref{fig:postdischargeRDA} show different perspectives of the bivariate dynamics of the COVID-19 effect in terms of the log hazard ratio on 30-day postdischarge readmission and death. The penalized likelihood approach was used to improve the smoothness of the estimated surface, with tuning parameters determined by the method of cross-validated deviance residuals. The two panels in the first row present 30-day postdischarge variations at 4 distinct dates of discharge. The downward sloping curves indicate that having COVID-19 was associated with significantly elevated risks of readmission and death, but only over the first week of discharge. The two panels in the second row present variations with calendar time on 4 different days after discharge, where the COVID-19 effect became less significant with more days since discharge. Within the first 5 days of discharge, the risk of readmission gradually increased as the pandemic unfolded, whereas the risk of death decreased until early June and then remained relatively unchanged afterward.

The remaining panels in the third and fourth rows of \Cref{fig:postdischargeRDA} are contour and surface plots, respectively, displaying the variations of the COVID-19-associated risks of readmission and death along two dimensions of time. Persistently declining log hazard ratios were observed from Day 0 to Day 30 since discharge, suggesting that the COVID-19 effects on readmission and death were decreasing with time. During the first 5 days after discharge, there existed three peaks for readmission around early April, early August, and mid October of 2020, while there was only one peak for death in early April. These findings are consistent with the evolution of the COVID-19 pandemic in the general population. In the initial phase of the pandemic, the case fatality rate was extremely high as the highly pathogenic variants of the novel coronavirus hit the country. Restricted access to health services and the fear of contagion contributed to deferred hospitalizations and readmissions, which supports the mildly high risk of readmission in early April. As governments implemented various mandates to contain the spread of the coronavirus, patients became more willing to be admitted to hospital, with the risk of readmission rebounding. In the meantime, the pervasive variants were getting less pathogenic, and hospitals became more prepared to treat COVID-19 patients, both of which led to a reduced case fatality rate. The risk of postdischarge death therefore decreased with calendar time.

In addition to modeling the bivariate COVID-19 effect on postdischarge outcomes, we tested its variation along two time dimensions according to \Cref{s:inferpen}. Consistent with the top right panel of \Cref{fig:postdischargeRDA}, the test of univariate variation across calendar time for postdischarge death led to a $p$-value of 0.727, indicating that the risk of death did not vary significantly with calendar time. All other tests of univariate and bivariate variation led to $p$-values less than 0.001.

\subsection{Discharge destinations}
In the second study, outcomes of interest were three options of discharge destination, including (1) in-hospital death or discharge to hospice, (2) discharge to a long- or short-term care hospital, skilled nursing facility, intermediate care facility, inpatient rehabilitation facility, psychiatric hospital, or critical access hospital (hereafter discharge to another facility), and (3) discharge to home with or without home care services, together viewed as mutually exclusive competing risks. Included in the data were 544,677 unplanned hospitalizations of 250,940 Medicare dialysis beneficiaries associated with 2,929 dialysis facilities throughout the year of 2020 (determined based on admission dates). Each hospital admission was followed up for up to 40 days. Among the 544,677 hospital admissions, 44,858 resulted in an in-hospital death or discharge to hospice; 125,723 were followed by a discharge to another facility; and 371,104 resulted in a home discharge. The remaining 2,992 admissions were associated with a hospital stay longer than 40 days, i.e., a censoring.

\begin{figure}
    \centering
    \begin{subfigure}[t]{0.9\textwidth}
        \centering
        \includegraphics[width=\linewidth]{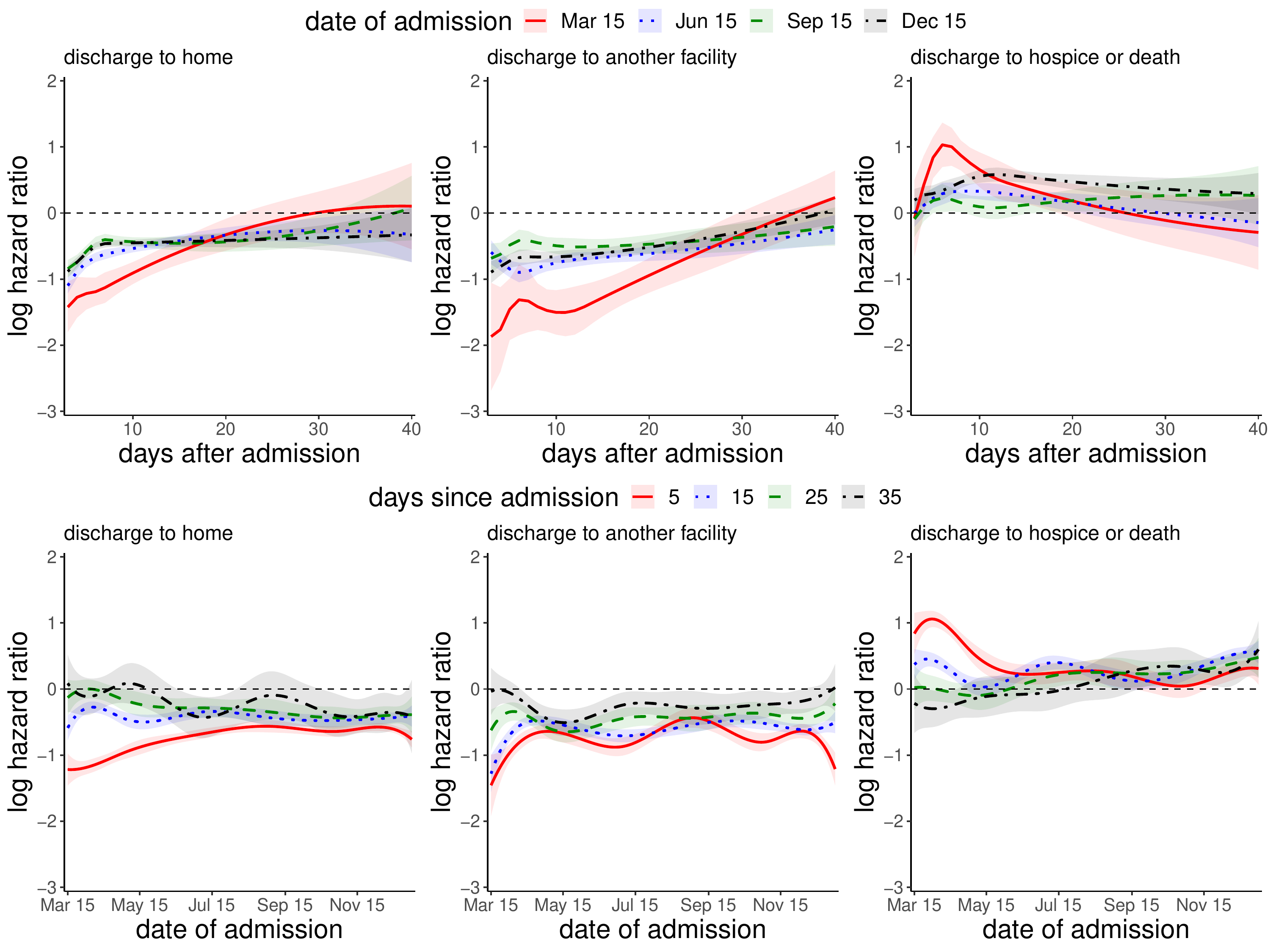} 
    \label{fig:discdestpostadcal}
    \end{subfigure}
    \begin{subfigure}[t]{0.3\textwidth}
        \centering
        \includegraphics[width=\linewidth]{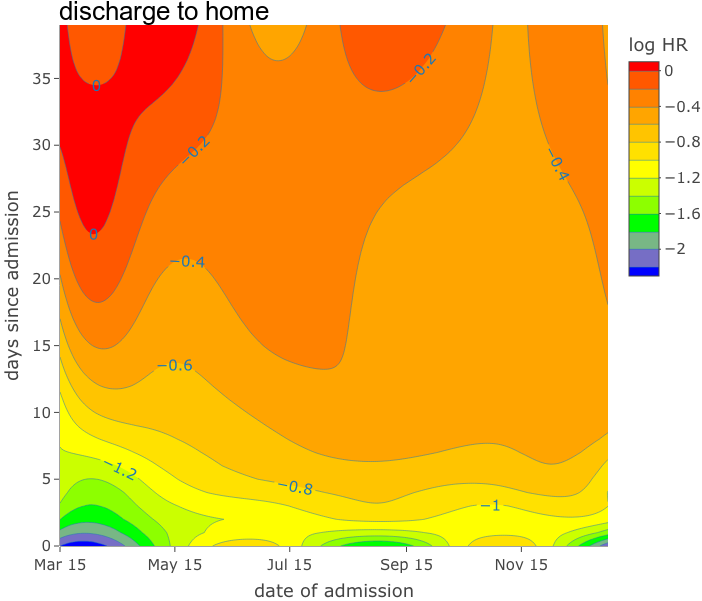} 
        \label{fig:disc2homecontour}
    \end{subfigure}
   \begin{subfigure}[t]{0.3\textwidth}
    \centering
        \includegraphics[width=\linewidth]{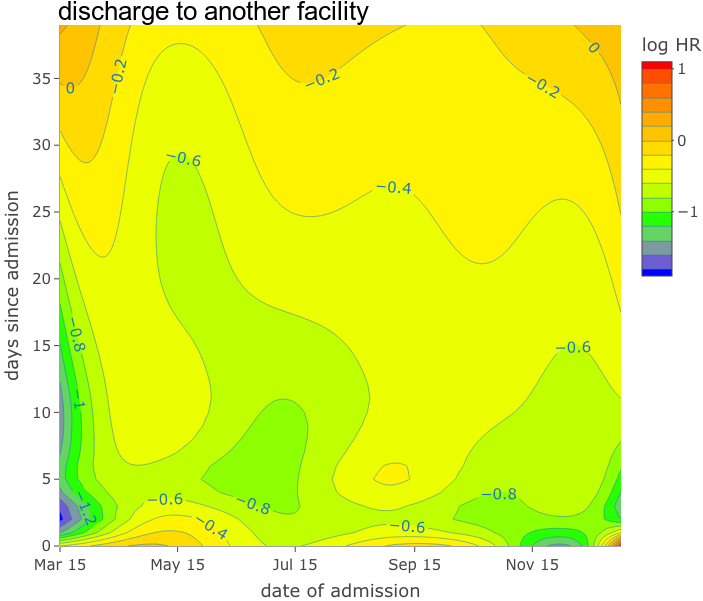}
        \label{fig:disc2faccontour}
    \end{subfigure}
    \begin{subfigure}[t]{0.3\textwidth}
    \centering
        \includegraphics[width=\linewidth]{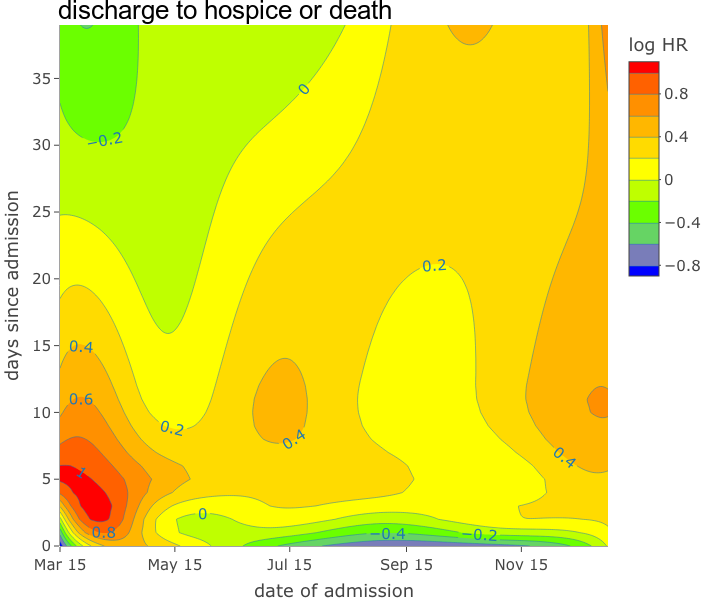} 
        \label{fig:disc2hospiceordeathcontour}
    \end{subfigure}
    \begin{subfigure}[t]{0.3\textwidth}
        \centering
        \includegraphics[width=\linewidth]{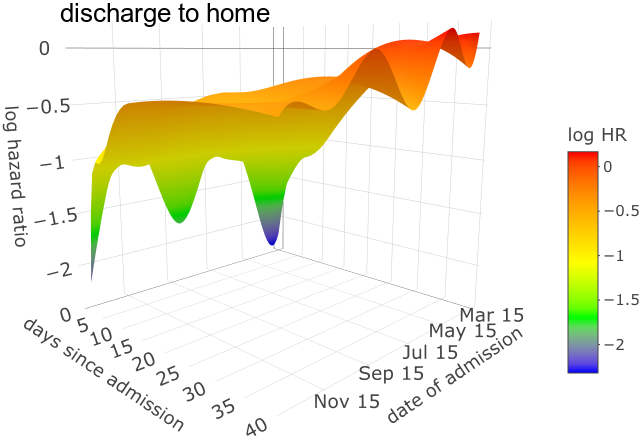} 
        \label{fig:disc2home3d}
    \end{subfigure}
   \begin{subfigure}[t]{0.3\textwidth}
    \centering
        \includegraphics[width=\linewidth]{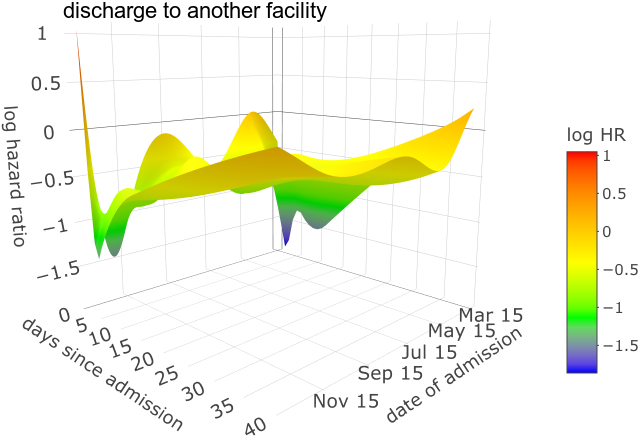}
        \label{fig:disc2fac3d}
    \end{subfigure}
    \begin{subfigure}[t]{0.3\textwidth}
    \centering
        \includegraphics[width=\linewidth]{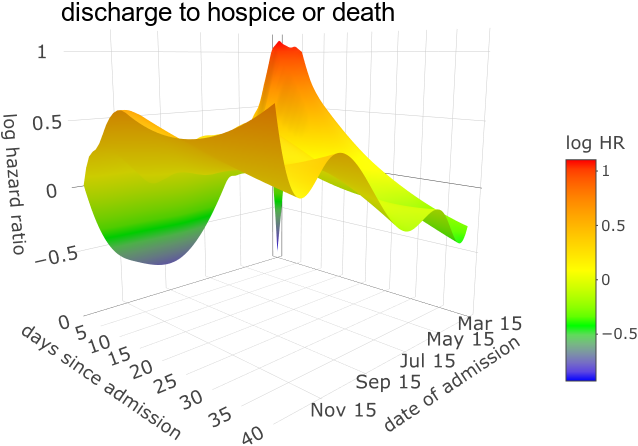} 
        \label{fig:disc2hospiceordeath3d}
    \end{subfigure}
    \caption{Bivariate variation of log hazard ratios with respect to in-hospital COVID-19 diagnosis for discharge status (home, another facility, and hospice/death). Included in the sample were 544,677 unplanned acute-care hospitalizations of 250,940 Medicare beneficiaries on dialysis associated with 2,929 Medicare-certified dialysis facilities in 2020. Ribbons in the top six panels indicate 95\% confidence intervals. Panels in the third and fourth rows are contour and surface plots, respectively.\label{fig:discdest}}
\end{figure}

We ran an unpenalized bivariate varying coefficient model to validate its performance on the discharge status data, in which the coefficient of COVID-19 was formulated as a bivariate function of post-admission time (i.e., days after admission or length of hospital stay) and calendar time. Similarly as before, the 12 panels of \Cref{fig:discdest} present the dynamics of the COVID-19 effect (in log hazard ratio) on three discharge destinations from different perspectives. Panels in the first two rows indicate that patients admitted with COVID-19 were less likely to be discharged to home or to another facility, and more likely to die in hospital or be discharged to hospice than those without COVID-19, especially in the initial phase of the pandemic and over the first 20 days of hospitalization. The COVID-19 effects remained significant with calendar time (the first row), but shrank as the length of stay increased (the second row). Evidence shown in the contour and surface plots (last two rows of \Cref{fig:discdest}) is consistent with what one would anticipate in the early stage of the pandemic: compared with those admitted without COVID-19, dialysis patients admitted with COVID-19 were associated with a significantly higher risk of early in-hospital death or discharge to hospice, and a significantly lower risk of early discharge to home or another facility. The COVID-19 effects then became less significant until mid-November 2020. After mid-November, the risk of in-hospital death or early discharge to hospice mildly increased among COVID-19 hospitalizations, while the risk of early discharge to another facility decreased substantially among COVID-19 hospitalizations, suggesting a worsening situation toward the end of 2020.

For all three discharge destinations, we performed tests of univariate and bivariate variation of the COVID-19 effects, similarly as in \Cref{s:postdischarge}. The resulting $p$-values were all less than 0.001, implying that the COVID-19 effects were significantly varying jointly or separately with post-admission and calendar time.

\section{Discussion}
\label{s:dis}
Motivated by our recent investigations into the dynamic impact of COVID-19 on dialysis patients, we have proposed a bivariate varying coefficient model for large-scale competing risks data. This novel model successfully characterizes the variation of COVID-19 effects on both event and calendar timescales. To address the computational challenge arising from fitting the model to the massive data in our applications, we developed an efficient tensor-product proximal Newton algorithm. Further, we introduced difference-based anisotropic penalization to alleviate model overfitting and the unsmoothness of the estimated effect surface. Various methods of cross-validation were considered for parameter tuning purposes. Statistical testing procedures with and without penalization were also designed to examine whether the COVID-10 effect variation was significant across event and calendar time, either jointly or separately. The proposed methods have been comprehensively evaluated through simulation studies and applications to dialysis patients amidst the COVID-19 pandemic.

Although inspired by COVID-19 studies on dialysis patients, the bivariate varying coefficient model can be harnessed in a variety of applications. For instance, among patients with breast cancer, evidence suggests that the racial and ethnic disparities in their cause-specific survival change significantly with post-diagnosis time \citep{wu2022scalable}. The proposed model can be leveraged to examine whether those disparities also change with age at diagnosis, thereby promoting health equity through more customized treatment options.

Multivariate varying coefficient models, as a flexible and granular analytical approach, have been studied in the presence of functional responses \citep{zhu2012multivariate, pietrosanu2021estimation} or longitudinal outcomes \citep{niu2019adjusting, wang2021kernel}. However, none of these studies has allowed the coefficients to depend on event time in a survival manner, which imposes a higher order of computational complexity than modeling event-time-independent varying coefficients in a multi-dimensional context. In contrast, our proposed model features bivariate effect dependence with both event time and an arbitrary risk factor; the accompanying inference, penalization, and model selection methods also advance the current literature of varying coefficient modeling.

\backmatter

\section*{Acknowledgements}
The authors thank Drs. Joseph M. Messana and Kirsten F. Herold (University of Michigan) for helpful discussion and comments, and Mr. Garrett W. Gremel (Optum, Inc.) for excellent data management. This work was supported by the Centers for Medicare and Medicaid Services (CMS, contract numbers 75FCMC18D0041 and 500-2016-00085C), the National Institute of Diabetes and Digestive and Kidney Diseases (NIDDK, project number R01DK129539), and the University of Michigan Institute for Computational Discovery \& Engineering Fellowship Program. The statements contained in this article are solely those of the authors and do not necessarily reflect the views or policies of the CMS and NIDDK.
%The authors assume responsibility for the accuracy and completeness of the information contained.

%  Not included in the original version of this paper!

\section*{Supporting Information}

%  Here, we create the bibliographic entries manually, following the
%  journal style.  If you use this method or use natbib, PLEASE PAY
%  CAREFUL ATTENTION TO THE BIBLIOGRAPHIC STYLE IN A RECENT ISSUE OF
%  THE JOURNAL AND FOLLOW IT!  Failure to follow stylistic conventions
%  just lengthens the time spend copyediting your paper and hence its
%  position in the publication queue should it be accepted.

%  We greatly prefer that you incorporate the references for your
%  article into the body of the article as we have done here 
%  (you can use natbib or not as you choose) than use BiBTeX,
%  so that your article is self-contained in one file.
%  If you do use BiBTeX, please use the .bst file that comes with 
%  the distribution.

\bibliography{ref.bib}

\section*{Data Availability Statement}
Because the data for dialysis patients contain protected health information and/or personally identifiable information, they will not be publicly available as required by the Centers for Medicare and Medicaid Services.

\appendix
\section{}

\subsection{Appendix A Gradient and Hessian of \texorpdfstring{$\ell_{jg}(\bgamma_j, \btheta_j)$}{text}}

For $g = 1, \ldots{}, G$, $i = 1, \ldots{}, n_g$, and $j = 1, \ldots{}, m$, we define
\[
S_{jgi}^{(u)}(\bgamma_j, \btheta_j, X_{gi}) 
\coloneqq \sum_{r \in R_g(X_{gi})} \exp\{\bL_{gr}^\top(X_{gi}) \bgamma_j + \bW_{gr}^\top \btheta_j\}
\begin{bmatrix}
\bL_{gr}(X_{gi}) \\
\bW_{gr}
\end{bmatrix}
^{\odot u}, \quad u = 0, 1, 2,
\]
where $\bL_{gr}(X_{gi}) \coloneqq \bZ_{gr} \otimes \brbB(\brX_{gr}) \otimes \bB(X_{gi})$, and for a vector $\bv \in \mathbb{R}^p$, $\bv^{\odot 0} \coloneqq 1$, $\bv^{\odot 1} \coloneqq \bv$, and $\bv^{\odot 2} \coloneqq \bv\bv^\top$. The gradient $\dot\ell_{jg}(\bgamma_j, \btheta_j)$ and Hessian $\ddot\ell_{jg}(\bgamma_j, \btheta_j)$ of $\ell_{jg}(\bgamma_j, \btheta_j)$ are hence given by
\begin{align}
\dot\ell_{jg}(\bgamma_j, \btheta_j)
&=\sum_{i=1}^{n_g} \Delta_{jgi} \left \{
\begin{bmatrix}
\bL_{gi}(X_{gi}) \\
\bW_{gi}
\end{bmatrix} - \bU^{(1)}_{jgi}(\bgamma_j, \btheta_j, X_{gi})\right\}, \label{eq:grad} \\
\ddot\ell_{jg}(\bgamma_j, \btheta_j)
&=-\sum_{i=1}^{n_g} \Delta_{jgi} \bV_{jgi}(\bgamma_j, \btheta_j, X_{gi}), \label{eq:hessian}
\end{align}
in which 
\begin{align*}
\bU^{(w)}_{jgi}(\bgamma_j, \btheta_j, X_{gi})
\coloneqq \frac{S_{jgi}^{(w)}(\bgamma_j, \btheta_j,  X_{gi})}{S_{jgi}^{(0)}(\bgamma_j, \btheta_j, X_{gi})}, \quad w = 1, 2, \\
\bV_{jgi}(\bgamma_j, \btheta_j, X_{gi})
\coloneqq \bU^{(2)}_{jgi}(\bgamma_j, \btheta_j, X_{gi}) - \{\bU^{(1)}_{jgi}(\bgamma_j, \btheta_j, X_{gi})\}^{\odot 2}.
\end{align*}

\subsection{Appendix B Proof of Proposition 1}

\begin{proof}
Let $\bM \coloneqq [\bC^{(t)}\bOmega_{jl}\{\bC^{(t)}\}^\top]^{-1}$, let $\bSigma$ denote the variance of $\bx \coloneqq \bC^{(t)} \{\vec(\tilde\bgamma^\top_{jl})-\tilde\bb_{jl}\}$, and let $Q \coloneqq \bx^\top \bM \bx$ denote the Wald test statistic. Since $\bSigma$ is orthogonally diagonalizable, there exists an orthogonal matrix $\bP$ such that $\bP \bSigma \bP^\top = \bPsi$, with $\bPsi$ being a diagonal matrix of positive eigenvalues of $\bSigma$. Let $\bR \coloneqq \bPsi^{-1/2}\bP $, a nonsingular matrix. Then $\bR \bSigma \bR^\top = \bI$. Since $(\bR^\top)^{-1} \bM \bR^{-1}$ is symmetric and orthogonally diagonalizable, there exists another orthogonal matrix $\bT$ such that $\bT(\bR^\top)^{-1} \bM \bR^{-1}\bT^\top = \bPhi$ is a diagonal matrix sharing the same eigenvalues $\mu_1, \ldots{}, \mu_{K\brK \times K\brK}$ as those of $(\bR^\top)^{-1} \bM \bR^{-1}$. Let $\bz \coloneqq \bT \bR \bx$. Then under the null $H^{(t)}_{0}$, we have $\bz \sim \mathcal{N}(\mathbf{0}, \bI)$. Since $\bT\bR$ is nonsingular, $\bx = \bR^{-1} \bT^\top \bz$. It follows that $Q = \bz^\top \bPhi \bz = \sum_{u=1}^{K\brK \times K\brK} \mu_u G^2_u$, where $G_u$'s independently follow the standard normal distribution. Observe that
\[
(\bR^\top \bT^\top)^{-1} \bM \bSigma \bR^\top \bT^\top = \bT(\bR^\top)^{-1} \bM \bSigma \bR^\top \bT^\top = \bT (\bR^\top)^{-1} \bM \bR^{-1} \bT^\top = \bPhi.
\]
This implies that $\bM \bSigma$ and $\bPhi$ have the same set of eigenvalues (since the mapping $\bA \mapsto \bB^{-1} \bA \bB$ preserves eigenvalues).
\end{proof}

\label{lastpage}

\subsection{Appendix C Proof of Proposition 2}

\begin{proof}
Given estimates $\hat\btheta_j$, $\hat\bbeta_j(\cdot, \cdot)$ for the bivariate varying coefficient model (1), the martingale residuals can be defined as
\[
\hat M_{jgi} \coloneqq \hat M_{jgi}(\infty, \brX_{gi}) = \Delta_{jgi} - \exp(\bW_{gi}^\top \hat\btheta_j)\int_{0}^{X_{gi}} \exp\left\{\bZ_{gi}^\top \hat\bbeta_j(t, \brX_{gi})\right\} \hat\lambda_{0jg}(t) \,\rd t,
\]
where the baseline hazard estimates $\hat\lambda_{0jg}(\cdot)$ are determined via the Breslow estimator. Further, the log-likelihood with respect to the $j$th failure type can be written as
\begin{equation*}
\begin{split}
& \sum_{g=1}^G \sum_{i=1}^{n_g} \{\Delta_{jgi} \log \lambda_{jgi}(X_{gi} \mid \bZ_{gi}, \bW_{gi}, \brX_{gi}) + \log S_{jgi}(X_{gi} \mid \bZ_{gi}, \bW_{gi}, \brX_{gi})\} \\
=&\, \sum_{g=1}^G \sum_{i=1}^{n_g}
\left[\Delta_{jgi} \{\bZ_{gi}^\top \bbeta_j(X_{gi}, \brX_{gi}) + \bW_{gi}^\top \btheta_j + \log \lambda_{0jg}(X_{gi}) \phantom{\int} \right. \\
&\qquad \left.-\int_0^{X_{gi}} \exp\{\bZ_{gi}^\top \bbeta_j(t, \brX_{gi}) + \bW_{gi}^\top \btheta_j\} \lambda_{0jg}(t)\, \rd t\right],
\end{split}
\end{equation*}
where $S_{jgi}(t \mid \bZ_{gi}, \bW_{gi}, \brX_{gi})$ is the corresponding survivor function. Assuming that the baseline hazard $\lambda_{0jg}(\cdot)$ is known, we have the deviance $D$ written as
\begin{equation*}
\begin{split}
\label{eq:deviance1}
D &= 2\sup_{\bbeta_{jgi}, \btheta_{jgi}}
\sum_{g=1}^G \sum_{i=1}^{n_g}
\left\{\Delta_{jgi} [\bZ_{gi}^\top \{\bbeta_{jgi} - \hat\bbeta_j(X_{gi}, \brX_{gi})\} + \bW_{gi}^\top (\btheta_{jgi} - \hat\btheta_j)] \phantom{\int} \right. \\
&\qquad \left.-\int_0^{X_{gi}} \left[\exp(\bZ_{gi}^\top \bbeta_{jgi} + \bW_{gi}^\top \btheta_{jgi}) - \exp\{\bZ_{gi}^\top\hat\bbeta_j(t, \brX_{gi}) + \bW_{gi}^\top\hat\btheta_j\}\right] \lambda_{0jg}(t)\, \rd t\right\},
\end{split}
\end{equation*}
where $\bbeta_{jgi}$ and $\btheta_{jgi}$ are subject-cause-specific estimates allowed in a saturated model. Now, we have the first order condition
\[
\Delta_{jgi} = \exp(\bZ_{gi}^\top \bbeta_{jgi} + \bW_{gi}^\top \btheta_{jgi}) \int_0^{X_{gi}}  \lambda_{0jg}(t)\, \rd t, \quad g = 1, \ldots{}, G,\, i = 1, \ldots{}, n_g.
\]
With this condition, the deviance $D$ reduces to
\begin{equation*}
\begin{split}
\label{eq:deviance2}
D &= -2 
\sum_{g=1}^G \sum_{i=1}^{n_g}
\left\{\Delta_{jgi} \log \frac{\exp\{\bZ_{gi}^\top\hat\bbeta_j(X_{gi}, \brX_{gi}) + \bW_{gi}^\top\hat\btheta_j\}\int_0^{X_{gi}}  \lambda_{0jg}(t)\, \rd t}{\Delta_{jgi}} + \tilde M_{jgi}\right\} \\
&= -2 
\sum_{g=1}^G \sum_{i=1}^{n_g}
\left[\Delta_{jgi} \left\{\bZ_{gi}^\top\hat\bbeta_j(X_{gi}, \brX_{gi}) + \bW_{gi}^\top\hat\btheta_j + \log\int_0^{X_{gi}}  \lambda_{0jg}(t)\, \rd t\right\} + \tilde M_{jgi}\right],
\end{split}
\end{equation*}
where 
\[
\tilde M_{jgi} \coloneqq \tilde M_{jgi}(\infty, \brX_{gi}) = \Delta_{jgi} - \exp(\bW_{gi}^\top \hat\btheta_j)\int_{0}^{X_{gi}} \exp\left\{\bZ_{gi}^\top \hat\bbeta_j(t, \brX_{gi})\right\} \lambda_{0jg}(t) \,\rd t
\]
is the martingale residual with known baseline hazard $\lambda_{0jg}(\cdot)$. Then the deviance residual $d_{jgi}$ for subject $i$ in the $g$th stratum with respect to the $j$th failure type can be written as
\[
d_{jgi} = \sign(\hat M_{jgi}) \sqrt{-2\left[\Delta_{jgi} \left\{\bZ_{gi}^\top\hat\bbeta_j(X_{gi}, \brX_{gi}) + \bW_{gi}^\top\hat\btheta_j + \log\int_0^{X_{gi}}  \hat\lambda_{0jg}(t)\, \rd t\right\} + \hat M_{jgi}\right]},
\]
where $\hat M_{jgi}$ is the martingale residual $\tilde M_{jgi}$ with $\lambda_{0jg}(\cdot)$ replaced by $\hat\lambda_{0jg}(\cdot)$.
\end{proof}

\subsection{Appendix D Supplementary Figure}

\begin{figure}[htbp]
\centering
\includegraphics[width=.9\textwidth]{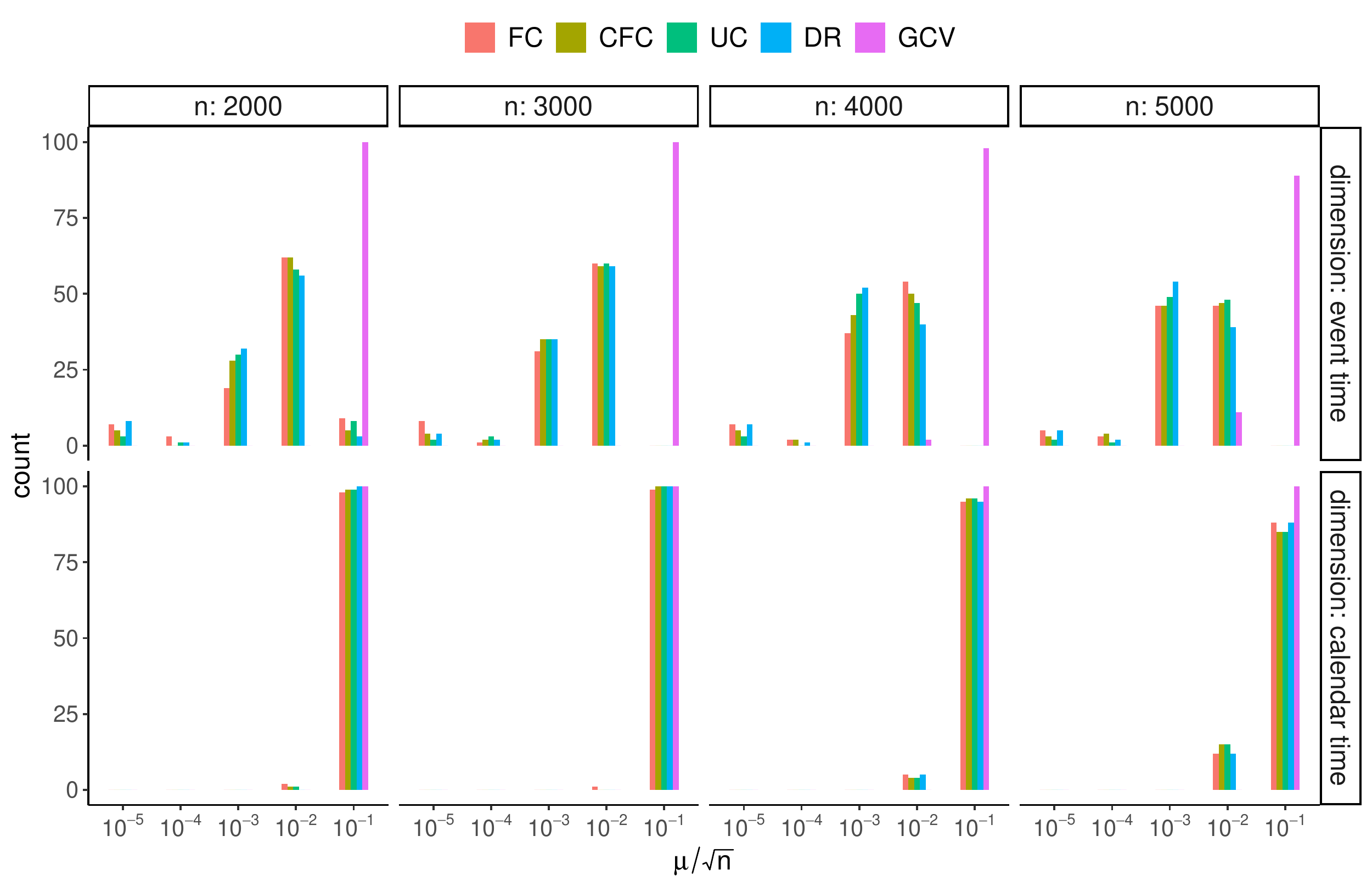}
\caption{A comparison of the distribution of selected tuning parameters for five cross-validation methods: fold-constrained (FC), complementary fold-constrained (CFC), and fold-unconstrained (UC) cross-validated partial likelihood, cross-validated deviance residuals (DR), and generalized cross-validation (GCV). In each scenario, 100 training and validation data replicates were generated independently. A 5-by-5 grid of tuning parameters was formed such that $\mu/\sqrt{n}$ (with $n$ denoting sample size) and $\brmu/\sqrt{n}$ varied from $10^{-5}$ to $10^{-1}$. Each cross-validation method was applied to a training data replicate to determine the optimal tuning parameters. True values were $\beta_1(t, \brx) = \sin(3\pi t/4)\exp(-0.5\brx)$ and $\beta_2(t, \brx) = 1$.}
\label{fig:CVtuningpara}
\end{figure}

\end{document}

%% file: preamble.tex
\usepackage{bm}
\usepackage{amssymb,mathtools}
\usepackage{natbib}
\usepackage{caption}
\usepackage{subcaption}
\usepackage[hyphens]{url}
\usepackage[linesnumbered, ruled]{algorithm2e}
\usepackage{hyperref}
\hypersetup{
colorlinks=true,
linkcolor=red,
filecolor=magenta,
citecolor=blue,
urlcolor=blue
}
\usepackage{cleveref}
\Crefname{assumption}{Assumption}{Assumptions}
\usepackage{setspace}
\usepackage{tikz}
\usepackage{calc} % for simple arithmetic
\tikzset{>=latex} % for LaTeX arrow head

\usepackage{bbm}
\usepackage{textcomp} % \textminus

\usepackage{multirow,multicol,makecell,booktabs}
\def\bbeta{\bm{\beta}}

\def\bgamma{\bm{\gamma}}

\def\beeta{\bm{\eta}}
\def\btheta{\bm{\theta}}

\def\bmu{\bm{\mu}}
\def\brbmu{\Breve{\bm{\mu}}}

\def\bSigma{\bm{\Sigma}}
\def\bGamma{\bm{\Gamma}}
\def\bOmega{\bm{\Omega}}
\def\bPsi{\bm{\Psi}}
\def\bPhi{\bm{\Phi}}

\def\diag{\mathrm{diag}}

\def\brmu{\Breve{\mu}}
\def\brd{\Breve{d}}
\def\brk{\Breve{k}}

\def\bru{\Breve{u}}
\def\brx{\Breve{x}}
\def\brB{\Breve{B}}

\def\brK{\Breve{K}}

\def\brX{\Breve{X}}

\def\brbB{\Breve{\mathbf{B}}}
\def\brbD{\Breve{\mathbf{D}}}
\def\brbI{\Breve{\mathbf{I}}}

\def\rd{\mathrm{d}}

\def\rF{\mathrm{F}}
\def\rM{\mathrm{M}}
\def\rP{\mathrm{P}}

\def\rS{\mathrm{S}}

\def\bA{\mathbf{A}}
\def\bB{\mathbf{B}}
\def\bC{\mathbf{C}}
\def\bD{\mathbf{D}}

\def\bI{\mathbf{I}}
\def\bL{\mathbf{L}}
\def\bM{\mathbf{M}}

\def\bP{\mathbf{P}}
\def\bQ{\mathbf{Q}}
\def\bR{\mathbf{R}}

\def\bT{\mathbf{T}}
\def\bU{\mathbf{U}}
\def\bV{\mathbf{V}}
\def\bW{\mathbf{W}}

\def\bZ{\mathbf{Z}}
\def\bb{\mathbf{b}}

\def\bx{\mathbf{x}}
\def\bv{\mathbf{v}}
\def\bz{\mathbf{z}}

\DeclareMathOperator{\sign}{sign}
\DeclareMathOperator{\tr}{trace}
\def\vec{\mathrm{vec}}
\def\CVE{\mathrm{CVE}}
\def\diag{\mathrm{diag}}